\documentclass[12pt]{iopart}

\usepackage{iopams,citesort,graphicx,tabularx}  

\begin{document}

% User defined commands
\newcommand{\rd}{\mathrm{d}}
\newcommand{\p}{\partial}
\newcommand{\px}{{\bi{x}}}
\newcommand{\expct}[1]{\langle #1 \rangle}
\newcommand{\Expct}[1]{\left\langle #1 \right\rangle}
\newcommand{\diff}[2]{\frac{\mathrm{d} #1}{\mathrm{d} #2}}
\newcommand{\prt}[2]{\frac{\partial #1}{\partial #2}}
\newcommand{\const}{\mathrm{const.}}
\renewcommand{\(}{\left(}
\renewcommand{\)}{\right)}
\renewcommand{\[}{\left[}
\renewcommand{\]}{\right]}
\newcommand{\im}{\mathrm{Im}}
\newcommand{\re}{\mathrm{Re}}

\title{Collective Lyapunov modes}

\author{Kazumasa A Takeuchi$^{1,2}$ and Hugues Chat\'e$^2$}

\address{$^1$ Department of Physics, the University of Tokyo,
 7-3-1 Hongo, Bunkyo-ku, Tokyo 113-0033, Japan.}
\address{$^2$ Service de Physique de l'\'Etat Condens\'e, CEA-Saclay, 91191 Gif-sur-Yvette, France.}
\ead{kat@kaztake.org}
\begin{abstract}
We show, using covariant Lyapunov vectors in addition to
 standard Lyapunov analysis,
 that there exists a set of \textit{collective Lyapunov modes}
 in large chaotic systems exhibiting collective dynamics. 
Associated with delocalized Lyapunov vectors,
 they act collectively on the trajectory and hence characterize the
 instability of its collective dynamics. 
We further develop, for globally-coupled systems,
a connection between these collective modes and the Lyapunov modes
 in the corresponding Perron-Frobenius equation. 
We thereby address the fundamental question of the effective dimension
 of collective dynamics and discuss
the extensivity of chaos in presence of collective dynamics.
\end{abstract}

%Uncomment for PACS numbers title message
% Keywords required only for MST, PB, PMB, PM, JOA, JOB? 
%\vspace{2pc}
%\noindent{\it Keywords}: Article preparation, IOP journals
% Uncomment for Submitted to journal title message
\submitto{\JPA}
% Comment out if separate title page not required
% \maketitle

\section{Introduction}

Emergence of collective behavior in large dynamical systems
 is a striking example of situations where microscopic interactions
 lead to non-trivial time-dependent macroscopic behavior, 
 in contrast with equilibrium systems:
 a collection of dynamical units is in general not self-averaging,
 and hence macroscopic observables may
% evolve periodically, quasi-periodically, or chaotically,
% even without exact synchronization of microscopic chaos
 show a variety of temporal evolutions
 even in the presence of microscopic chaos and in the absence of synchronization
 \cite{Chate.Manneville-PTP1992,Kaneko-PRL1990,Pikovsky.Kurths-PRL1994,Kaneko-PD1995}.
Such non-trivial collective behavior (NTCB) is now known to be quite generic,
 being observed whether interactions are local or global,
 time variable is discrete or continuous
 \cite{Hakim.Rappel-PRA1992,Nakagawa.Kuramoto-PTP1993,Nakagawa.Kuramoto-PD1995,Brunnet.etal-PD1994},
 and evolution is deterministic or noisy
 \cite{Shibata.etal-PRL1999,DeMonte.etal-PRL2004,DeMonte.etal-PTPS2006}.
One of the most interesting features of NTCB
 is that microscopic chaos can coexist with macroscopic evolutions
 of different instability -- periodic, quasi-periodic, and even chaotic
 in the case of global coupling \cite{Bohr.etal-PRL1987,Kaneko-PD1995}.
It is therefore natural to ask whether emerging macroscopic behavior
 can be captured by some of the Lyapunov exponents (LEs),
 which measure the infinitesimal rates of exponential divergence
 of nearby trajectories in phase space \cite{Eckmann.Ruelle-RMP1985}.
If such \textit{collective} Lyapunov modes exist,
 they should shed light not only on
 how macroscopic and microscopic instabilities coexist
 in the single Lyapunov spectrum of the system,
 but also on a number of fundamental questions.
For instance, the set of the collective LEs
 would give the effective dimension of the collective behavior
 and allow us to study its analogy to small dynamical systems,
 which had only been inferred so far from the observed dynamics.
Furthermore, collective Lyapunov modes also call for redefining
 extensivity of chaos \cite{Ruelle-CMP1982},
 which is usually taken to be the existence of a well-defined LE density
 per unit volume, numerically examined by measuring Lyapunov spectra
 at different system sizes and collapsing them after rescaling
 of the exponent index by the system size.
Since NTCB is well-defined in the infinite-size limit,
 collective LEs characterizing NTCB are expected to be intensive,
 so that the above definition of the extensivity should not hold as it stands.
Indeed, in contrast to spatiotemporal chaos in one spatial dimension,
 for which extensivity has been numerically shown in various generic models
 (see, e.g., \cite{Manneville-LNP1985,Keefe-PLA1989,Livi.etal-JPA1986}),
 in higher dimensions where NTCB can take place,
 only few studies could suggest extensivity
 \cite{OHern.etal-PRE1996,Egolf.etal-N2000}
 within rather narrow ranges of system sizes.

Given this importance,
 a few earlier studies attempted to capture collective Lyapunov modes,
 without, however, reaching a definitive answer.
Shibata and Kaneko \cite{Shibata.Kaneko-PRL1998}
 and, independently, Cencini \textit{et al.} \cite{Cencini.etal-PD1999}
 argued that one needs to study \textit{finite-amplitude} perturbations
 to quantify the instability of collective chaos in globally-coupled maps
 (see \cite{Cencini.Vulpiani-JPA2012} for a review on such finite-size LEs).
Showing that macroscopic instability is captured
 when the amplitude of the perturbation exceeds $\mathcal{O}(1/\sqrt{N})$,
 where $N$ is the system size,
 they implied that the standard LEs for infinitesimal perturbations
 do not reflect the collective dynamics.
On the other hand, Nakagawa and Kuramoto
 studied a collective chaos regime of globally-coupled limit-cycle oscillators
 and showed that Lyapunov spectra at different system sizes
 overlapped onto each other, if a few LEs
 at both ends of the spectra
 were taken away before the spectra were rescaled by the system size
 \cite{Nakagawa.Kuramoto-PD1995}.
They speculated that these LEs are associated with the collective dynamics,
 but they had to find such collective LEs by trial and error
 and, more importantly, could not provide any criterion to define them.
Indeed, we and coworkers have recently shown that,
% at the two ends of the spectrum of globally coupled systems, in general,
 for globally-coupled systems, in general,
 there are \textit{subextensive} bands of LEs 
 that should be scaled by $\log N$ instead of $N$
 at both ends of the spectrum,
 which sandwich the extensive LEs in between \cite{Takeuchi.etal-PRL2011}.
Therefore, Nakagawa and Kuramoto may have actually taken away
 such subextensive LEs from the spectrum to obtain the remaining branch
 of extensive LEs, within the system sizes they studied,
 instead of truly collective Lyapunov modes.

This frustrating situation was overcome when
 Ginelli \textit{et al.} proposed an efficient algorithm
 to compute covariant Lyapunov vectors (CLVs) in large dynamical systems
 \cite{Ginelli.etal-PRL2007,Ginelli.etal-JPA2012}.
The CLVs are the vectors spanning the subspaces of the Oseledec decomposition
 of tangent space \cite{Eckmann.Ruelle-RMP1985}.
Each LE $\lambda^{(j)}$ has its associated CLV $\delta\px^{(j)}$
 at any point on the trajectory $\px(t)$%
\footnote{
Similarly, in this article,
 $\delta(\cdot)^{(j)}$ denotes the infinitesimal change
 in the quantity $(\cdot)$ caused by the infinitesimal perturbation
 $\delta\px^{(j)}$ to the trajectory $\px$ along the $j$th CLV.
},
 which provides the intrinsic direction of perturbation
 growing at the rate $\lambda^{(j)}$.
Therefore, the vector components of the CLVs indicate
 which dynamical units constitute the corresponding Lyapunov modes.
Using this property of the CLVs, we, in our preceding work
 \cite{Takeuchi.etal-PRL2009}, reported numerical evidence
 that collective behavior of large dynamical systems is indeed
 encoded in their Lyapunov spectrum:
 while most modes are localized on a few degrees of freedom,
 thus corresponding to microscopic fluctuations of the system,
 there exist some \textit{collective modes} characterized
 by the delocalized CLVs, acting therefore collectively on the dynamical units.
The delocalization of the CLVs is quantified
 by the mean value of the inverse participation ratio (IPR)
 \cite{Mirlin-PR2000}
\begin{equation}
 Y_2^{(j)} \equiv \Bigl\langle{\sum_i |\delta x_i^{(j)}|^4}\Bigr\rangle_t,  \label{eq:Y2Def}
\end{equation}
 where $\delta x_i^{(j)}$ is the $i$th component of the CLV $\delta\px^{(j)}$
 normalized with the $L^2$-norm $\sum_i |\delta x_i^{(j)}|^2 = 1$
 and the brackets $\expct{\cdots}_t$ indicate
 the average taken along the trajectory.
Given that $Y_2$ provides the inverse of the average number
 of degrees of freedom participating in the mode $\delta\px^{(j)}$,
 we expect $Y_2 \sim 1/N$ for the collective modes,
 as the number of the participating components
 increases with the system size $N$,
 while $Y_2$ stays constant for large $N$
 for the remaining microscopic modes.
This is our definition of the collective and microscopic modes,
 or the delocalized and localized modes,
 and provides a clear criterion for distinguishing them.
% collective and microscopic Lyapunov modes,
In this way we indeed identified a few collective modes
 in collective chaos exhibited by globally-coupled limit-cycle oscillators
 \cite{Takeuchi.etal-PRL2009}.
We also showed, for globally-coupled maps with additive noise,
 that the collective modes in the standard Lyapunov analysis
 also arise as Lyapunov modes for the corresponding Perron-Frobenius dynamics,
 further justifying our definition of the collective modes.

This approach is further developed in the present article.
After providing a detailed description on the detection
 of the collective modes in the limit-cycle oscillators,
 we shall study the role of each collective mode
 in the observed macroscopic dynamics (Section \ref{sec:Col}).
Then, taking another, rather simple example of collective chaos
 in globally-coupled noisy maps,
 we show a quantitative correspondence between
 collective modes and Perron-Frobenius Lyapunov modes
 and discuss the generality of the role of the collective modes
 found in the preceding section (Section \ref{sec:ColPF}).
On this basis, in Section \ref{sec:ColChaos}, we study
 a regime of truly non-trivial collective chaos
 and show the interplay of macroscopic instabilities in this NTCB regime.
We shall deal in particular with the effective dimension of collective chaos,
 revisiting earlier work by Shibata \textit{et al.} \cite{Shibata.etal-PRL1999}
 which concluded that the dimension $D$ increases as $D \sim -\log\sigma$
 with decreasing noise amplitude $\sigma$ and hence
 the dimension is infinite in the noiseless limit $\sigma\to 0$.
Here, we show how this dimension is related to the collective dynamics
 and to the macroscopic instabilities, and discuss the generality of our results.
%question, though not definitively,
% their conclusion of the infinite dimensionality in the noiseless limit.
Section \ref{sec:Dis} is devoted to discussions and concluding remarks.

\section{Existence of collective Lyapunov modes}  \label{sec:Col}

We first demonstrate that the standard Lyapunov spectrum does contain
 collective Lyapunov modes,
 without the need to invoke finite-amplitude perturbations.
Following
 Nakagawa and Kuramoto's work
 \cite{Nakagawa.Kuramoto-PTP1993,Nakagawa.Kuramoto-PD1995} and
 our preceding letter \cite{Takeuchi.etal-PRL2009},
 we consider here $N$ globally-coupled limit-cycle oscillators
\begin{equation}
 \dot{W}_i = W_i - (1+\mathrm{i}c_2)|W_i|^2 W_i + K (1+\mathrm{i}c_1) (\expct{W} - W_i),  \label{eq:GLDef}
\end{equation}
 with $i=1,2,\cdots, N$, complex variables $W_i$,
 and the global field $\expct{W} \equiv \frac{1}{N}\sum_i W_i$.
The parameter values are set to be $c_1 = -2.0, c_2 = 3.0, K = 0.47$,
 which correspond to a regime of collective chaos \cite{Nakagawa.Kuramoto-PTP1993,Nakagawa.Kuramoto-PD1995,Takeuchi.etal-PRL2009}.
We numerically integrate Equation \eref{eq:GLDef} by the fourth-order
 Runge-Kutta method with time step $0.1$
 and compute LEs and CLVs using Ginelli \textit{et al.}'s algorithm 
 \cite{Ginelli.etal-PRL2007,Ginelli.etal-JPA2012}.
Most data presented in this section are recorded
 over a period longer than $10^5$
 after a transient of length $10^4$ or more is discarded.
When full Lyapunov spectrum is computed,
 we allow a transient period of $100N$ or more
 for the convergence of the Lyapunov exponents and vectors.

\begin{figure}[t]
 \begin{center}
  \includegraphics[clip]{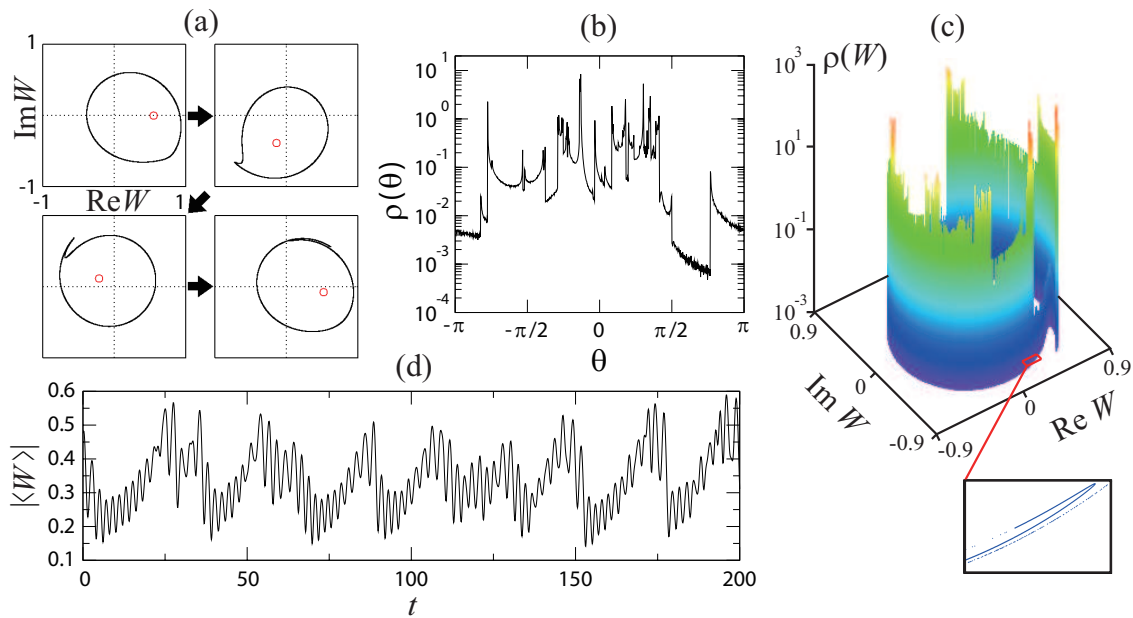}
  \caption{Collective chaos in $N=10^7$ globally-coupled limit-cycle oscillators \eref{eq:GLDef}. (a) Series of snapshots of the oscillators (dots) in the complex plane, taken at a fixed time interval $0.8$. The red circles indicate the position of the global field $\expct{W}$. (b) Typical instantaneous distribution for the angular position $\theta_i \equiv \arg W_i$ of the oscillators, here corresponding to the first snapshot in the panel (a). (c) Instantaneous distribution of the oscillators in the complex plane, at a moment near folding of the supporting line. The bottom panel shows the positions of the oscillators within the region indicated in the main panel. (d) Time series of the modulus of the global field, $|\expct{W}|$.}
  \label{fig:2-1}
 \end{center}
\end{figure}%

Collective chaos exhibited by these limit-cycle oscillators is shown
 in Figure \ref{fig:2-1}.
Because of the coupling to the global field,
 individual oscillators do not follow the limit-cycle oscillation
 but rotate rather irregularly with varying amplitudes and angular velocities,
 leading to stretching and folding of the supporting line
 (Figure \ref{fig:2-1}(a)).
This makes the local density of oscillators quite intricate
 with many peaks and fractal-like multilayer structure
 (Figure \ref{fig:2-1}(b,c)).
%, which is reminiscent of the invariant measure of single chaotic maps.
This imbalanced, fluctuating distribution of the oscillators
 drives irregular behavior of the global field $\expct{W}$
 and other macroscopic variables,
 which, in the studied regime,
 takes the form of a weak chaotic modulation of a quasiperiodic signal
 as seen in the time series of the modulus, $|\expct{W}|$
 (Figure \ref{fig:2-1}(d)).
This evolution resembles chaos emerging from a quasiperiodic regime in small systems
such as in two coupled nonlinear oscillators
 with incommensurate frequencies \cite{Sano.Sawada-PLA1983}.

\begin{figure}[t]
 \begin{center}
  \includegraphics[clip,width=.9\hsize]{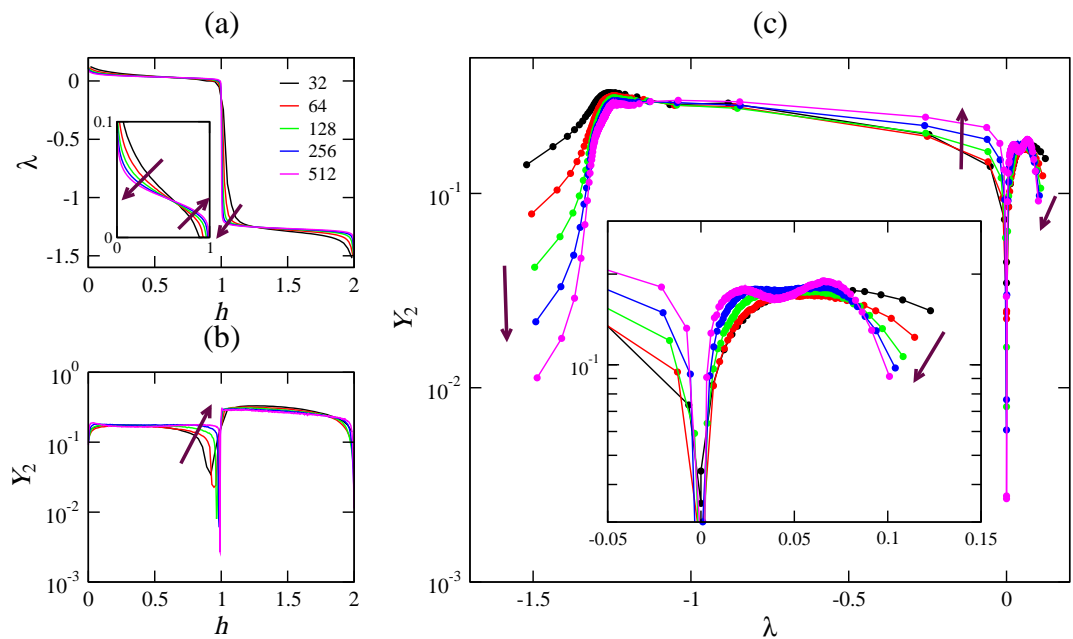}
  \caption{Spectra of LEs $\lambda^{(j)}$ against the rescaled index $h \equiv (j-0.5)/N$ (a), those of IPRs $Y_2^{(j)}$ (b), and the parametric plots $(\lambda^{(j)},Y_2^{(j)})$ (c), measured at different sizes $N$ for the globally-coupled limit-cycle oscillators \eref{eq:GLDef}. The arrows indicate increasing $N$. The insets show close-ups of the positive branch ($\lambda^{(j)} > 0$) of the spectra.}
  \label{fig:2-2}
 \end{center}
\end{figure}%

The values of the LEs $\lambda^{(j)}$ and the IPRs $Y_2^{(j)}$
 in this system are shown in Figure \ref{fig:2-2}(a,b)
 against the rescaled index $h \equiv (j-0.5)/N$.
Reflecting the LEs of an uncoupled limit-cycle oscillator,
 one zero and one negative, the Lyapunov spectra of the coupled system
 have two branches, near $0.05$ and $-1.25$ in this case
 (Figure \ref{fig:2-2}(a)).
The spectra at different sizes, however, do not collapse
 onto a single curve, but show rather strong size dependence (inset).
This indicates that the system is not entirely extensive
 as expected for globally-coupled systems \cite{Takeuchi.etal-PRL2011}.
The spectra of the IPRs also tend to stay around some constant values
 within the two branches, indicating that most CLVs are localized
 according to our definition, but systematic drifts are also visible
 near both ends of the two branches (Figure \ref{fig:2-2}(b)).
The situation is better presented when the IPRs $Y_2^{(j)}$
 are plotted against the corresponding LEs $\lambda^{(j)}$,
 as shown in Figure \ref{fig:2-2}(c).
Here we clearly see that $Y_2^{(j)}$ decreases with increasing $N$
 at both ends of the spectra as well as near $\lambda^{(j)}=0$,
 while most Lyapunov modes are concentrated in the regions
 where the values of $Y_2^{(j)}$ hardly depend on $N$.
This suggests that the Lyapunov spectra may contain
 delocalized, collective modes near the largest, smallest, and zero LEs,
 but not elsewhere for the system sizes studied here.

\begin{figure}[t]
 \begin{center}
  \includegraphics[clip,width=.95\hsize]{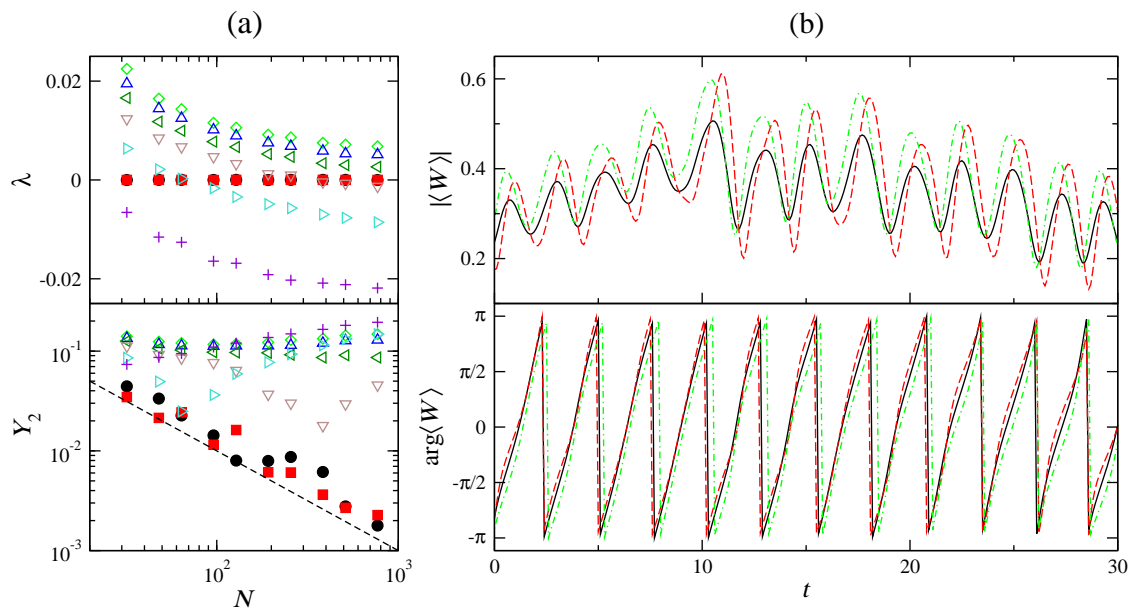}
  \caption{Collective modes at $\lambda^{(j)} = 0$ in globally-coupled limit-cycle oscillators \eref{eq:GLDef}. (a) LEs $\lambda^{(j)}$ and IPRs $Y_2^{(j)}$ as functions of the system size $N$. The dashed line indicates $Y_2 \sim 1/N$. The two neutral modes (solid symbols) are delocalized and thus collective, whereas other modes nearby (open symbols) are microscopic. (b) Time-series of the modulus and the argument of the global field $\expct{W}$ at $N=256$, for the unperturbed trajectory $\bi{W}$ (black solid line) as well as for the ``perturbed ones'' $\bi{W}+c_1\delta\bi{W}^{\mathrm{(zero1)}}$ (red dashed line) and $\bi{W}+c_2\delta\bi{W}^{\mathrm{(zero2)}}$ (greed dot-dashed line) with arbitrary constants $c_1$ and $c_2$. Here the former and the latter shift the phase for $|\expct{W}|$ and $\arg\expct{W}$, respectively, though in general these degenerated Lyapunov modes are arbitrary superpositions of the two fundamental perturbations.}
  \label{fig:2-3}
 \end{center}
\end{figure}%

Now we demonstrate that
 there indeed exist a few collective modes in these regions.
Concerning the region near $\lambda^{(j)} = 0$,
 the system has two neutral modes associated with invariance
 under time translation and angular rotation.
They are delocalized and thus collective
 as their $Y_2^{(j)}$ decreases as $1/N$
 (solid symbols in Figure \ref{fig:2-3}(a)).
Indeed, these collective modes shift the two phases
 of the quasiperiodic collective dynamics under the chaotic modulation,
 as shown in time series of the global field for the unperturbed trajectory
 $\bi{W}$ and for those shifted in the direction of the two corresponding CLVs,
 $\delta\bi{W}^{\mathrm{(zero1)}}$ and $\delta\bi{W}^{\mathrm{(zero2)}}$
 (Figure \ref{fig:2-3}(b)).
In contrast, other nearby LEs change their values smoothly as $N$ is varied,
 some of them crossing the zero line
 (open symbols in Figure \ref{fig:2-3}(a)).
The IPRs of these modes are essentially independent of $N$,
 except when their LEs become accidentally close to zero,
 in which case the numerical scheme cannot resolve well
 the degeneracy with the collective zero modes.
This clearly indicates that these near-zero modes are microscopic.

\begin{figure}[t]
 \begin{center}
  \includegraphics[clip,width=.95\hsize]{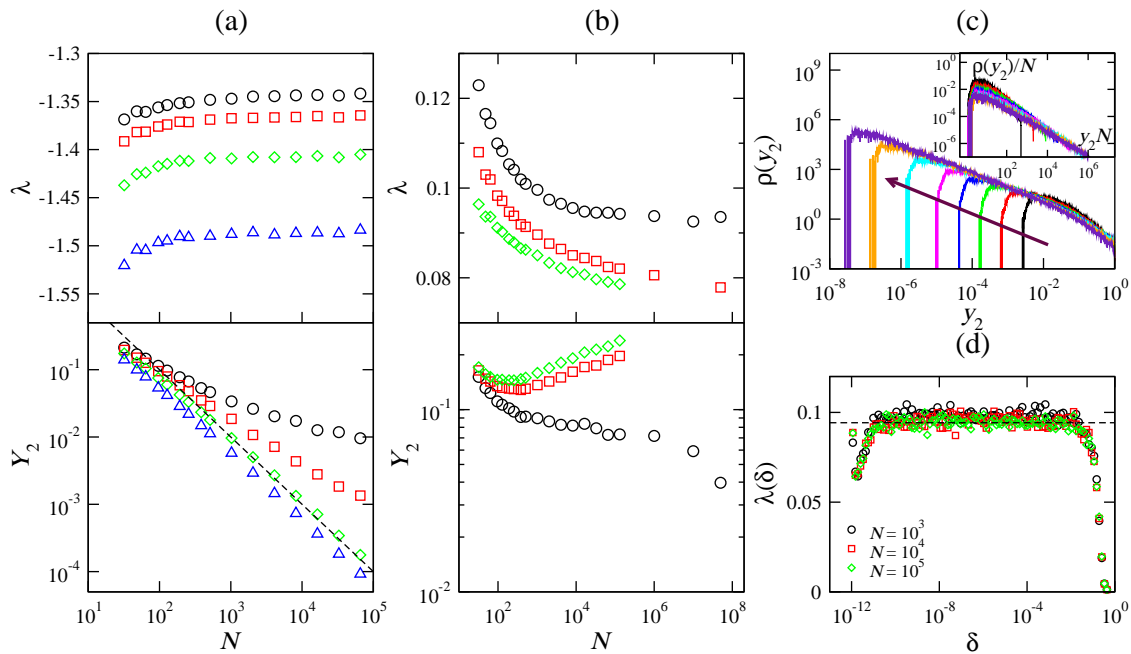}
  \caption{First and last Lyapunov modes in globally-coupled limit-cycle oscillators \eref{eq:GLDef}. (a,b) LEs $\lambda^{(j)}$ and IPRs $Y_2^{(j)}$ against the system size $N$, for the last four modes (a) and the first three modes (b). The dashed line in the panel (a) indicates $Y_2 \sim 1/N$. (c) Histogram of the instantaneous IPR $y_2$ for the first mode, measured at $N = 512, 2048, \cdots, 131072, 10^6, 10^7, 5 \times 10^7$ (increasing as indicated by the arrow). The abscissa is multiplied by $N$ in the inset. (d) Finite-size LE $\lambda(\delta)$ as a function of the amplitude $\delta$ of the perturbation. The dashed line shows the value of the largest LE $\lambda^{(1)}$ at $N=131072$.}
  \label{fig:2-4}
 \end{center}
\end{figure}%

For the negative end of the spectrum,
 Figure \ref{fig:2-4}(a) shows the values of the LEs $\lambda^{(j)}$
 and the IPRs $Y_2^{(j)}$
 as functions of the system size $N$ for the last four modes.
Although their IPR values decrease more slowly than $1/N$
 for small system sizes, when $N$ is increased further%
\footnote{
For globally-coupled systems, one can explicitly write down
 the \textit{backward} evolution for tangent space dynamics,
 using stored information of trajectory.
This allows us to compute LEs (and CLVs) from both ends of the spectrum,
 so that we can study the last Lyapunov modes
 even at very large system sizes.},
 the last two modes become completely delocalized, i.e., $Y_2^{(j)} \sim 1/N$,
 whereas the IPRs of the other modes decrease more and more slowly
 towards some asymptotic non-zero values.
Therefore, only the last two Lyapunov modes are collective here.
The situation is similar at the positive end of the spectrum
 (Figure \ref{fig:2-4}(b)),
 except that one needs to explore extremely large system sizes:
 the IPR of the first mode starts to decrease faster and faster
 only above $N=10^6$, and is still far from the $1/N$ decay
 at $N = 5 \times 10^7$.
However, the histogram of the instantaneous values of the IPR,
 $y_2^{(1)} \equiv \sum_i |\delta W_i^{(1)}|^4$,
 reveals that the distribution function actually scales as $1/N$,
 except the upper bound which is fixed at $y_2^{(1)} = 1$ by definition
 (Figure \ref{fig:2-4}(c)).
This indicates the asymptotic scaling of $Y_2^{(1)} \sim 1/N$
 and therefore the first Lyapunov mode is collective.
This conclusion can also be confirmed by using finite-amplitude perturbations.
Shibata and Kaneko \cite{Shibata.Kaneko-PRL1998}
 and Cencini \textit{et al.} \cite{Cencini.etal-PD1999}
 showed, somewhat speculatively, that the mean expansion rate $\lambda(\delta)$
 of perturbations of size $\delta$,
 called the finite-size LE \cite{Cencini.Vulpiani-JPA2012},
 are equal to the largest LE $\lambda^{(1)}$
 for $\delta \ll \delta_{\rm c} \sim 1/N$
 and to the largest expansion rate of the collective dynamics
 for $\delta \gg \delta_{\rm c}$.
In our case, since the first mode is collective,
 the finite-size LE $\lambda(\delta)$ exhibits a single plateau
 at the value of the first LE $\lambda^{(1)}$
 regardless of the system size $N$ (Figure \ref{fig:2-4}(d)).

\begin{figure}[t]
% \begin{center}
  \includegraphics[clip,width=\hsize]{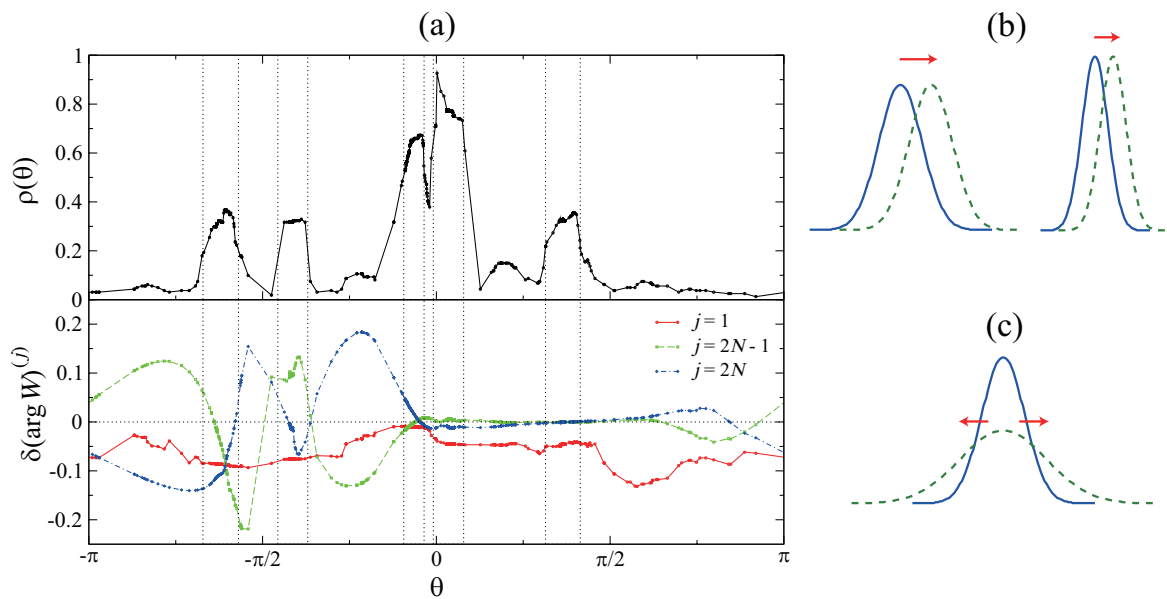}
  \caption{Role of the collective modes with positive and negative LEs ($j=1$ and $j = 2N-1$ and $2N$, respectively) in the globally-coupled limit-cycle oscillators \eref{eq:GLDef} at $N=512$. (a) Instantaneous profile for the density $\rho(\theta)$ of the oscillators (upper panel) and the shift in their angular positions induced by the collective Lyapunov modes, $\delta(\arg W_i)^{(j)}$ (lower panel). The vertical dotted lines indicate the regions of the five main peaks in the density profile. The local density $\rho(\theta_i)$ is measured from the number of oscillators within the range $[\theta_i - \pi/20, \theta_i+\pi/20]$. The profile of $\delta(\arg W_i)^{(1)}$ is locally averaged with the same window, while the raw data are shown for $\delta(\arg W_i)^{(2N-1)}$ and $\delta(\arg W_i)^{(2N)}$. (b,c) Schematic drawings illustrating the role of the positive (b) and negative (c) collective modes.}
  \label{fig:2-5}
% \end{center}
\end{figure}%

The role of these three collective Lyapunov modes can also be studied
 from the structure of the corresponding CLVs.
Although, strictly, the full structure of the CLVs
 as well as of the oscillator density in the complex plane
 should be considered, we have confirmed that
 the components of the CLVs for the collective modes are mostly
 along the supporting line of the oscillators,
 so that the vector structures can be faithfully represented
 on the angular coordinates, $\theta_i \equiv \arg W_i$.
Figure \ref{fig:2-5}(a) shows an instantaneous profile
 of the oscillator density $\rho(\theta)$
 projected on the angular coordinates $\theta_i$ (upper panel)
 and the shift in the angular positions induced
 by each of the three collective modes with positive and negative LEs,
 $\delta(\arg W_i)^{(j)} \equiv \lim_{\epsilon \to 0} [\arg(W_i + \epsilon\delta W_i^{(j)}) - \arg W_i]/\epsilon = \im(\delta W_i^{(j)}/W_i^{(j)})$ (lower panel).
One can see that the shift due to the positive collective mode,
 $\delta(\arg W_i)^{(1)}$ (red solid line),
 is varying along the angular coordinates,
 except in the peaked regions of the oscillator density
 (indicated by vertical dotted lines)
 where the values of $\delta(\arg W_i)^{(1)}$ stay practically constant.
% entirely negative
% (or positive, since the global sign of the CLVs is arbitrary)
% with varying amplitudes.
%The amplitudes 
This indicates that the positive collective mode moves
 these peaks in the density profile, almost uniformly,
 with different amplitudes (Figure \ref{fig:2-5}(b)).
Since such a shift makes a direct perturbation to the global field $\expct{W}$,
 which behaves chaotically in the present regime,
 this mode is naturally assigned with the positive LE $\lambda^{(1)}$.
By contrast, for the two negative collective modes,
 the angular shifts $\delta(\arg W_i)^{(2N-1)}$ and $\delta(\arg W_i)^{(2N)}$
 vary widely and often change their sign within the peaked regions
 (green dashed and blue dot-dashed lines).
The two tails of the peaks are then moved in the opposite directions,
 and hence the peak widths are expanded or narrowed (Figure \ref{fig:2-5}(c)).
Since these modes have negative LEs,
% it implies that
 such perturbations leading to changes in the peak widths
 decay exponentially fast.
In other words, these negative collective modes tend to adjust
 the width of the peaks, maintaining local synchronization of the oscillators.
The two neutral collective modes have already been studied
 in Figure \ref{fig:2-3}(b) and turned out to shift
 the two fundamental phases of the remaining quasiperiodic behavior.

To summarize, we have found
 one positive, two neutral, and two negative collective modes,
 at least up to the largest sizes we studied.
This set of the collective LEs is analogous to small dynamical systems
 exhibiting bifurcation from quasiperiodic to chaotic behavior
 \cite{Sano.Sawada-PLA1983}, but here
 the associated CLVs act on collections of oscillators,
 controlling dynamics of peaks in the density profile.
We have shown, therefore, that collective Lyapunov modes do exist
 within the framework of the standard Lyapunov analysis,
 characterized by the delocalization of the CLVs,
 without the need to rely on finite-amplitude perturbations.
Unlike the finite-amplitude perturbations
 which capture the largest expansion rate in the collective dynamics,
 the collective Lyapunov modes characterize, we believe,
 full instability that emerges at the collective level.
However, we have also noticed that we need to overcome
 strong finite-size effects, in order for collective modes
 to be clearly decoupled from neighboring microscopic modes
 (see Figure \ref{fig:2-4}(a,b)),
 which is unfortunately quite demanding with the current machine power.
The situation becomes even worse for locally coupled systems
 such as coupled map lattices,
 for which one needs at least two spatial dimensions to study NTCB
 \cite{Chate.Manneville-PTP1992}
 and, further, local correlation of dynamical units
 reduces the effective system size,
 though we believe that collective Lyapunov modes
 should exist in locally-coupled systems as well.
In addition, for globally-coupled systems,
 the existence of collective chaos also helps us to capture
 collective Lyapunov modes,
 since for large system sizes
 one can typically probe only first few Lyapunov modes
 with positive LEs.
Moreover, globally-coupled systems allow us to investigate
 the infinite-size limit ``directly'' from the evolution
 of the instantaneous density profile,
 which is given by the nonlinear Perron-Frobenius (PF) equation
 \cite{Pikovsky.Kurths-PRL1994,Shibata.etal-PRL1999,Kaneko-PD1995}.
Lyapunov exponents measured in the PF dynamics
 have therefore been related to the collective dynamics \cite{Kaneko-PD1995},
 without, however, direct evidence being ever shown.
In the next section,
 following our preceding letter \cite{Takeuchi.etal-PRL2009},
 we will show that there is indeed a quantitative correspondence
 between the collective Lyapunov modes and some of such PF Lyapunov modes.

\section{Correspondence to Perron-Frobenius Lyapunov modes}  \label{sec:ColPF}

The PF equation describes the evolution
 of the instantaneous distribution function of the dynamical variables,
 denoted here by $\rho^t(x)$.
This idea can be applied to a single dynamical unit
 for globally-coupled systems,
 since the global field is given by $\rho^t(x)$,
 which makes the PF equation nonlinear
 \cite{Pikovsky.Kurths-PRL1994,Shibata.etal-PRL1999,Kaneko-PD1995}.
Then, one
% cannot solve the equation analytically in most cases but
 needs to rely on numerical approaches in most cases,
 typically discretizing the support of the evolving distribution
 into sufficiently fine bins.
It is unfortunately unrealistic for
 the globally-coupled limit-cycle oscillators
 studied above,
 because of the intricate, fractal-like structure of the support
 in the complex plane
 (see Figure \ref{fig:2-1}(c)).
To avoid this difficulty, we study here
 a simpler case of globally-coupled maps of a real variable,
 whose support is typically a bounded interval of the real axis.
%We also add noise to the system to avoid singularity
% in the corresponding PF equation, as explained below.

Specifically, we consider the following system:
\begin{equation}
 x_i^{t+1} = (1-K)f(x_i^t) + K \expct{f(x)} + \xi_i^t,  \label{eq:GCMDef}
\end{equation}
 with the logistic map $f(x) = 1-ax^2$
 and an iid noise $\xi_i^t$,
 which is added here to avoid singularities in the PF dynamics
 as explained below.
In the infinite-size limit $N \to \infty$,
 the corresponding nonlinear PF equation reads
\begin{equation}
 \rho^{t+1}(x) = \int \rho_{\rm N}(F^t(y) - x) \rho^t(y) \rd y,  \label{eq:PFDef1}
\end{equation}
 where $\rho_{\rm N}(\xi)$ is the probability density of the noise and
\begin{equation}
 F^t(y) = (1-K)f(y) + K \int f(z)\rho^t(z) \rd z.  \label{eq:PFDef2}
\end{equation}
If the system is noiseless, i.e., if $\xi_i^t = 0$,
% the delta function would replace $\rho_{\rm N}(\xi)$
% and induce singularity in the density profile $\rho^t(x)$,
% which is difficult to deal with in numerical simulations.
 $\rho_{\rm N}(\xi)$ is replaced by the delta function
 and hence
\begin{equation}
 \rho^{t+1}(x) = \frac{1}{1-K} \sum_\textrm{\scriptsize $y$ s.t. $F^t(y) = x$} \frac{\rho^t(y)}{|f'(y)|}, \quad (\textrm{if noiseless}),  \label{eq:PFNoiseless}
\end{equation}
 where the summation is taken over the preimages of $x$
 and $f'(y)$ is the first derivative of $f(y)$.
Following these preimages and dealing with the singularity
 due to the superstable point $f'(y) = 0$
 make the system practically inaccessible by numerical means.
This singularity is tamed by adding noise to the system
 as in Equation \eref{eq:GCMDef},
%Adding noise is a common remedy for this,
 or by studying a heterogeneous system with site-dependent
 local parameters \cite{Shibata.Kaneko-PRL1998,Cencini.etal-PD1999},
 which do not destroy collective dynamics
 but provides a useful situation
 to elucidate the nature of the collective behavior
 \cite{Shibata.etal-PRL1999,DeMonte.etal-PRL2004,DeMonte.etal-PTPS2006}.
%In particular, Shibata \textit{et al.} computed
% LEs for the PF equation \eref{eq:PFDef1} and showed that
% the Kaplan-Yorke dimension $D$ grows with decreasing noise amplitude $\sigma$
% as $D \sim -\log\sigma$, suggesting infinite dimensionality
% of collective chaos in the noiseless limit \cite{Shibata.etal-PRL1999}.

Computation of the PF equation \eref{eq:PFDef1} and its tangent space dynamics
 is straightforward, but has a few pitfalls
 which have been overlooked by some earlier studies.
% including the one by Shibata \textit{et al.}
First, the noise distribution $\rho_{\rm N}(\xi)$ must be bounded
 in such a way that the distribution of the dynamical units, $\rho^t(x)$,
 is always confined within the basin of attraction.
Gaussian noise cannot be used here for this reason.
Second, because the tangent space dynamics reads
\begin{equation}
 \fl \delta\rho^{t+1}(x) = \int \rd y \[ \rho_{\rm N}(F^t(y) - x) \delta\rho^t(y) + K\rho'_{\rm N}(F^t(y) - x)\rho^t(y) \int f(z)\delta\rho^t(z) \rd z \]  \label{eq:PFTangentSpace}
\end{equation}
 and involves the derivative of $\rho_{\rm N}(\xi)$,
 the noise distribution must be differentiable
 for proper computation of LEs and CLVs.
%Moreover, we will see in the following that
% $\rho_{\rm N}(\xi)$ should actually be differentiable up to higher orders
% to compute more than one Lyapunov modes reliably.
For these reasons, we choose the Kumaraswamy noise distribution
 \cite{Jones-SM2009}
 $\rho_{\rm N}(\xi) = \alpha\beta(\xi')^{\alpha-1}(1-(\xi')^\alpha)^{\beta-1}$
 with $\xi' \equiv (\xi/\sigma+1)/2 \in [0,1]$, $\alpha=3$, and $\beta=5$,
 for which the probability density is unimodal,
 bounded within $\xi \in [-\sigma,\sigma]$,
 twice differentiable, and nearly symmetric.

We start with a rather simple regime of collective chaos
 under strong coupling,
 where dynamical variables tend to synchronize
 to the chaotic dynamics of the uncoupled logistic map,
 but are weakly scattered by microscopic chaos and noise
 \cite{DeMonte.etal-PRL2004,DeMonte.etal-PTPS2006,Teramae.Kuramoto-PRE2001}.
Specifically, we set the local logistic parameter at $a=1.57$,
 which corresponds to the one-band chaos regime,
 and in the present section $K = 0.28$ and $\sigma = 0.1$,
 unless otherwise specified.
We perform both direct simulations of the maps \eref{eq:GCMDef} with finite $N$
 and simulations of the PF equation \eref{eq:PFDef1},
 which should correspond to the $N\to\infty$ limit.
It is indeed checked by confirming that
 the evolution of $\rho^t(x)$ from the PF equation
 matches that of large collections of the maps for many time steps,
 when starting from almost identical initial density profiles.
In the following, we compare Lyapunov modes obtained by the two methods.
Most data are recorded over $3\times 10^5$ time steps
 after a transient period of $300N$ or $50000$ steps for the maps,
 and recorded over $10^5$ steps after discarding $10^4$ steps
 for the PF equation,
 simulated with sufficiently many bins of equal width
 (typically 1024 bins within the interval $[-1.0, 1.5]$ in this section).

\begin{figure}[t]
 \begin{center}
  \includegraphics[clip,width=.95\hsize]{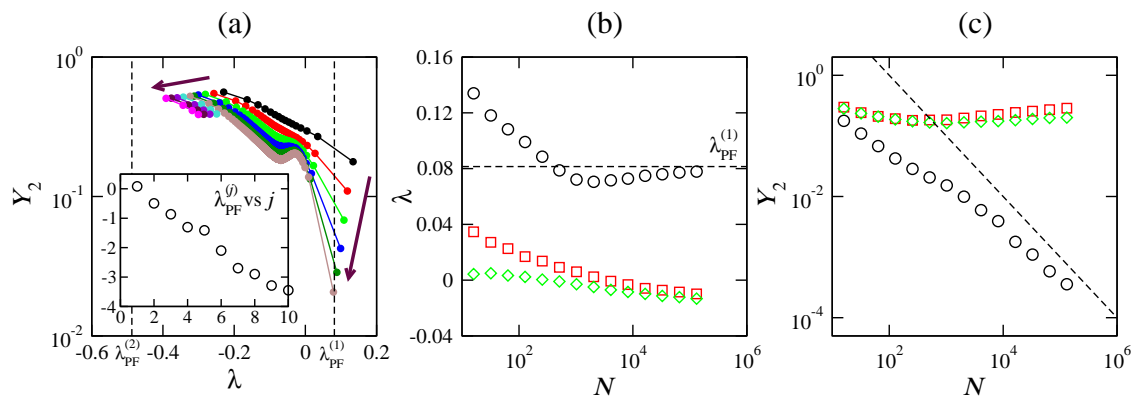}
  \caption{Lyapunov analysis for globally-coupled noisy logistic maps \eref{eq:GCMDef} with $a=1.57, K=0.28, \sigma = 0.1$ and the corresponding PF equation \eref{eq:PFDef1}. (a) IPR $Y_2^{(j)}$ against LE $\lambda^{(j)}$ for the maps with $N = 16, 32, \cdots, 512$. The last four modes are shown additionally for $N = 2048, 8192, \cdots, 131072$. The arrows indicate increasing $N$. The inset shows the spectrum of LEs for the PF dynamics, $\lambda_{\rm PF}^{(j)}$. The values of the first and second PF LEs, namely $\lambda_{\rm PF}^{(1)} = 0.081$ and $\lambda_{\rm PF}^{(2)} = -0.49$, are indicated by the vertical dashed lines in the main panel. (b,c) First three LEs (b) and their IPRs (c) as functions of the size $N$ of the maps. The dashed line in the panel (b) indicates the value of the first PF LE $\lambda_{\rm PF}^{(1)}$ and that in the panel (c) indicates $Y_2 \sim 1/N$.}
  \label{fig:3-1}
 \end{center}
\end{figure}%

Figure \ref{fig:3-1}(a) displays
 the full parametric plots of $(\lambda^{(j)}, Y_2^{(j)})$
 for the maps up to $N=512$,
 showing signs of delocalization for first few modes.
Increasing the system size further,
 we find that only the first one is actually delocalized, i.e.,
 $Y_2^{(1)} \sim 1/N$,
 while the following ones are localized (Figure \ref{fig:3-1}(c)).
The values of the LEs also vary differently for the first mode,
 which remains well separated from the others
 irrespective of the size $N$ (Figure \ref{fig:3-1}(b)).
Moreover, comparison to the LEs for the PF equation
 reveals that this first, collective LE tends to the first PF LE,
 suggesting $\lim_{N\to\infty} \lambda^{(1)} = \lambda_{\rm PF}^{(1)}$.
For the following Lyapunov modes, however, 
 we do not find such direct correspondence
 within the system sizes probed here.
The PF LEs $\lambda_{\rm PF}^{(j)}$ decrease so rapidly
 with increasing index $j$ (Figure \ref{fig:3-1}(a) inset)
 that even the second one $\lambda_{\rm PF}^{(2)}$
 is smaller than the last LE of the maps, $\lambda^{(N)}$
 (main panel).
The values of the last few LEs decrease as the system size increases,
 but they do not reach $\lambda_{\rm PF}^{(2)}$
 nor indicate any sign of delocalization in the associated IPRs
 up to the largest size we examined, namely $N=131072$.
The first Lyapunov mode is therefore the only collective mode
 present at the studied system sizes,
 so that it takes the same LE value as the first PF Lyapunov mode.

\begin{figure}[t]
 \begin{center}
  \includegraphics[clip,width=.9\hsize]{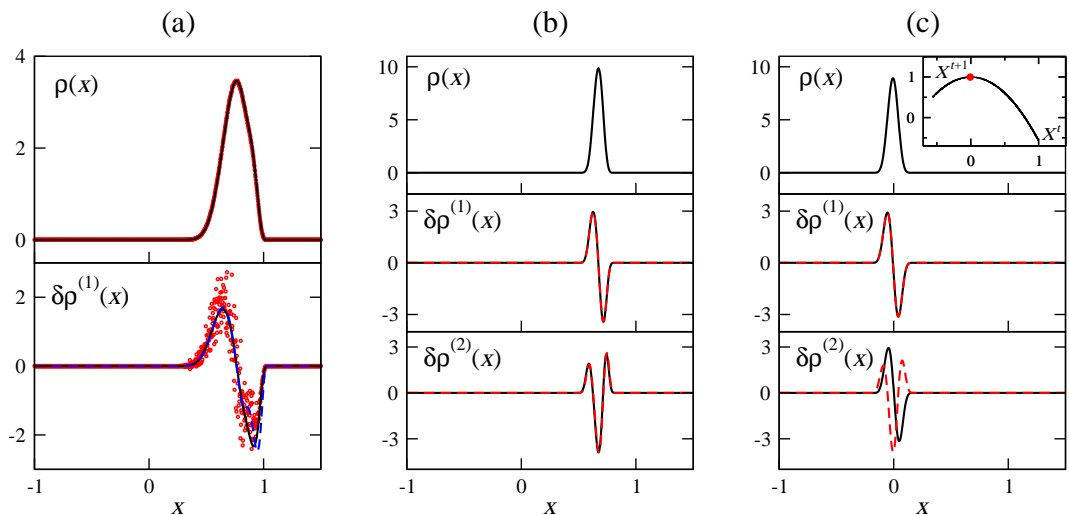}
  \caption{Snapshots of the distribution $\rho(x)$ and CLVs $\delta\rho^{(j)}(x)$ for globally-coupled noisy logistic maps \eref{eq:GCMDef} with $a=1.57$. (a) Typical snapshots for $K=0.28$ and $\sigma = 0.1$. Black solid lines and red circles indicate the results from PF and the maps of size $N=10^7$, respectively. The blue dashed line shows the (normalized) first derivative of $\rho(x)$, which overlaps the PF CLV $\delta\rho^{(j)}(x)$ for the most part. (b,c) Snapshots for $K=0.70$ and $\sigma = 0.1$, taken at a typical moment (b) or when the global field $X^t \equiv \expct{x^t}$ is near the superstable point (c). The black solid lines indicate the results from PF, whereas the red dashed lines show the $j$th derivative of $\rho(x)$. The inset of the panel (c) shows the return map of the global field, the red point indicating the moment when the snapshots are taken.}
  \label{fig:3-2}
 \end{center}
\end{figure}%

The correspondence between this collective mode
 and the first PF Lyapunov mode is underpinned
 from the structure of their CLV.
Figure \ref{fig:3-2}(a) compares the distribution $\rho(x)$ (top panel)
 and the first CLV $\delta\rho^{(1)}(x)$ (bottom panel)
 obtained directly from the PF equation (black solid line)
 and those constructed from the maps of size $N=10^7$ (red circles),
 at a moment when the two distributions coincide very well.
One can see that the first CLV from the maps exerts the same shift
 $\delta\rho^{(1)}(x)$ on the distribution as that from the PF equation
 (bottom panel).
It indicates that this collective mode,
 identified in the original system by the standard Lyapunov analysis,
 should indeed converge to the PF Lyapunov mode in the infinite-size limit.
Moreover, we notice in Figure \ref{fig:3-2}(a)
 that the spatial profile of the CLV $\delta\rho^{(1)}(x)$ is very close
 to the first derivative of the distribution, $\p_x\rho(x)$
 (blue dashed line), which is normalized here
 with the same $L^2$-norm as the CLV.
Since $\rho(x+\Delta x) \simeq \rho(x) + \p_x\rho(x)\Delta x$
 for small $\Delta x$, this collective mode with the positive LE
 merely shifts the position of the peak,
 similarly to the positive collective mode
 found in the globally-coupled limit-cycle oscillators
 (see Figure \ref{fig:2-5}).

The similarity to the derivative is more prominent
 for larger coupling strengths $K$.
For $K=0.70$, the first PF CLV $\delta\rho_{\rm PF}^{(1)}(x)$
 and the first derivative $\p_x\rho(x)$ are almost indistinguishable,
 and, actually, so are the $j$th PF CLV $\delta\rho_{\rm PF}^{(j)}(x)$
 and the $j$th derivative $\p_x^j\rho(x)$ for most times
 (Figure \ref{fig:3-2}(b)), unless $j$ is too large.
In particular, the second PF CLV $\delta\rho_{\rm PF}^{(2)}(x)$ resembles
 the second derivative $\p_x^2\rho(x)$,
 which can be regarded as the ``diffusion term'' that changes the width
 of the peak.
The associated LE being negative, $\lambda_{\rm PF}^{(2)} = -1.69$ here,
 this second PF Lyapunov mode maintains a constant width of the peak
 for each moment of the evolution.
This is again similar to the negative collective modes
 in the limit-cycle oscillators, though the second PF mode
 in the present system is not captured by the Lyapunov analysis of the maps
 at least for the studied system sizes.
The similarity of the $j$th CLV to the $j$th derivative
 is, however, lost occasionally, when the global field
 $X^t \equiv \expct{x^t}$ visits the close vicinity of the superstable
 point in its return map (Figure \ref{fig:3-2}(c); recall that the global field
 evolves like a single uncoupled logistic map in this case).
Then, since $F^t(y)$ in Equation \eref{eq:PFDef2} hardly depends on $y$,
 Equation \eref{eq:PFDef1} merely yields
 $\rho^{t+1}(x) \approx \rho_{\rm N}(1-x)$ irrespective of $\rho^t(x)$.
This results in the near degeneracy of all the CLVs,
 as indicated in Figure \ref{fig:3-2}(c).

In this section, we have shown a quantitative correspondence
 between a collective Lyapunov mode identified for the maps
 and a Lyapunov mode of the corresponding PF dynamics,
 both having the same value of the LE and applying the same perturbation
 to the collection of the dynamical units in the infinite-size limit.
The correspondence has been demonstrated
 for the (single) positive Lyapunov mode in the strong-coupling regime
 of the globally-coupled noisy logistic maps,
 whereas we could not capture a PF Lyapunov mode with negative LE
 by analyzing the maps,
 for the system sizes probed in the present study.
In fact, we consider that negative Lyapunov modes in the PF representation
%, of which there are infinitely many,
 may not necessarily be present in the original dynamical systems.
A trivial example is an uncoupled collection of identical chaotic maps,
 e.g., at $K=0$ for the current system \eref{eq:GCMDef},
 where all Lyapunov modes are degenerate at the single positive LE,
 while the PF dynamics yields infinitely many negative LEs.
These negative PF modes should correspond to
 superpositions of local perturbations to the local maps,
 so that the perturbation grows exponentially in phase space
 but the perturbed distribution relaxes towards the invariant measure.
In contrast, we believe that PF modes with zero or positive LE
 should be all present as collective modes in the original system,
 because they cannot be superpositions
 of such incoherent microscopic perturbations.
Similarly, we conjecture that \textit{all} collective modes
 in dynamical systems should also arise in the corresponding PF equations,
 regardless of the sign of the LEs,
 because our definition of the collective modes indicates that
 $\mathcal{O}(N)$ dynamical units are perturbed thereby,
 which should also have impacts on the distribution.
A question that may naturally arise here
 is whether one can distinguish such ``physical'' PF Lyapunov modes
 from ``spurious'' ones,
 which are missing in the original dynamical systems,
 among infinitely many negative PF LEs.
Hyperbolicity of the Lyapunov modes could be a key to tackle this problem,
 as in spatially-extended dissipative systems where
 the Lyapunov spectrum consists of
 a finite number of mutually connected physical modes
 and the remaining spurious modes
 that are hyperbolically decoupled from all the physical modes
 \cite{Yang.etal-PRL2009,Takeuchi.etal-PRE2011},
 but first attempts in this direction worked only for the trivial case
 of the fully synchronized collective chaos (data not shown).
Elucidating this possible decoupling of the physical and spurious
 PF Lyapunov modes for general cases of NTCB
 is a fundamental issue left for future studies.

\section{Dimension and nature of non-trivial collective chaos}  \label{sec:ColChaos}

%The collective behavior studied in the present section
The study in the previous section
 allowed us to establish the direct connection between
 a collective mode and a PF Lyapunov mode,
% in a dynamical system and a Lyapunov mode in the corresponding PF equation,
 but we dealt with a rather trivial regime of collective behavior
 with strong coupling, which reduces for $K=0.70$
 to the fully synchronized state of the local maps in the noiseless limit.
Now we turn our attention to a regime of non-trivial collective chaos,
 where dynamical units loosely form a number of coevolving clusters,
 mutually connected and interacting
 through chaotic behavior of the global field.
We focus in particular on a weak-coupling regime
 of the globally-coupled noisy logistic maps \eref{eq:GCMDef}
 with $a=1.86$ and $K=0.10$,
 studied earlier by Shibata \textit{et al.} \cite{Shibata.etal-PRL1999}.
They measured the Kaplan-Yorke dimension $D_{\rm KY}$ from the PF LEs
 and found it to grow with decreasing noise amplitude $\sigma$ as
 $D_{\rm KY} \sim -\log\sigma$, using however the Gaussian noise distribution,
 for which one cannot numerically evolve the PF equation \eref{eq:PFDef1}
 in a proper way.
Here, using the controlled Kumaraswamy noise with various amplitudes,
 we revisit this problem of the effective dimension and, moreover,
 examine the instabilities of collective chaos
 on the basis of the Lyapunov approach developed in the previous sections.
%Unfortunately, in such a weakly coupled NTCB regime,
% it is practically impossible to reach collective modes
% by direct simulations of the maps, because they should be buried
% amidst the overwhelming number of microscopic modes
% located far from both ends of the spectrum.
%We therefore investigate mainly the PF Lyapunov modes here
% and show how they act on the dynamically evolving distribution of the units.
The transient and recording periods are typically
 $300N$ and $3 \times 10^5$ time steps for the maps
 and $10^4$ and $10^5$ time steps for the PF equation,
 respectively.

\begin{figure}[t]
 \begin{center}
  \includegraphics[clip,width=.9\hsize]{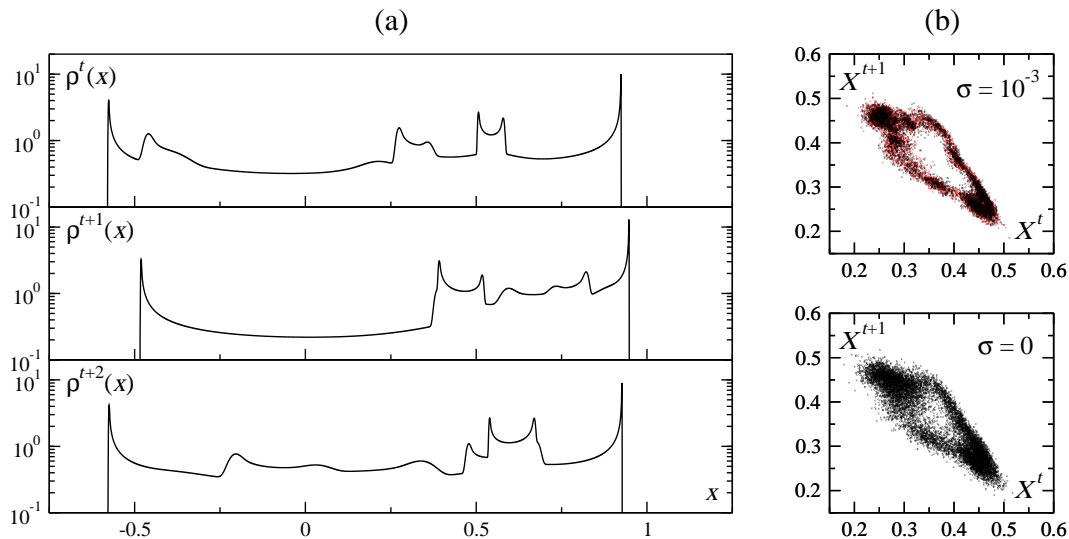}
  \caption{Collective chaos dynamics in globally-coupled noisy logistic maps \eref{eq:GCMDef} with $a=1.86$ and $K=0.10$. (a) Three consecutive snapshots of the instantaneous distribution $\rho^t(x)$ for $\sigma = 1 \times 10^{-3}$, obtained from the evolution of the PF equation \eref{eq:PFDef1} with $65536$ bins. (b) Return maps of the global field $X^t \equiv \expct{x^t}$ for $\sigma = 1 \times 10^{-3}$ (top) and $\sigma = 0$ (bottom). The black and red dots indicate the results from $10^7$ maps and the PF equation, respectively.}
  \label{fig:4-1}
 \end{center}
\end{figure}%

Figure \ref{fig:4-1}(a) shows the
 time evolution of the instantaneous distribution $\rho^t(x)$ in this regime,
 with noise amplitude $\sigma = 1 \times 10^{-3}$.
The distribution now consists of a number of peaks,
 connected to each other by gradually varying low-density regions,
 in striking contrast to the strong-coupling regime (Figure \ref{fig:3-2}(a))
 but rather similarly to the collective chaos in the limit-cycle oscillators
 (Figure \ref{fig:2-1}(b)).
Here, on top of the broad support,
 the rightmost, sharpest peak is created
 from the superstable point of $F^{t-1}(y)$,
 i.e., $y=0$ (see Equation \eref{eq:PFDef2});
 its image would be singular if no noise were added,
 but in the presence of noise it is scattered
 and leaves a peak of finite width, roughly $\sigma$.
This peak is then transported to different positions on each time step,
 with its width broadened by noise and chaos,
 and is finally dispersed over the support.
Because of this intricate structure of the distribution,
 the return map of the global field $X^t \equiv \expct{x^t}$
 is not simple and single-valued any more (Figure \ref{fig:4-1}(b) top panel),
 being scattered around the mean values.
This readily suggests that
 the effective dimension of the collective chaos
 is not as low as in the strong-coupling case.
Somewhat counterintuitively,
 eliminating the noise does not reduce this scattering but, instead,
 enhances it (Figure \ref{fig:4-1}(b) bottom panel),
 because the noise smooths the intricate structure
 of the distribution $\rho^t(x)$ and simplifies the collective dynamics.
The noise also allows us to simulate the evolution of $\rho^t(x)$ precisely
 by the PF equation with finite numbers of bins:
 for $\sigma = 1 \times 10^{-3}$, for example, a large collection of maps
 and the PF equation yield statistically indistinguishable results
 for the return map (Figure \ref{fig:4-1}(b) top panel).

\begin{figure}[t]
 %\begin{center}
  \includegraphics[clip,width=\hsize]{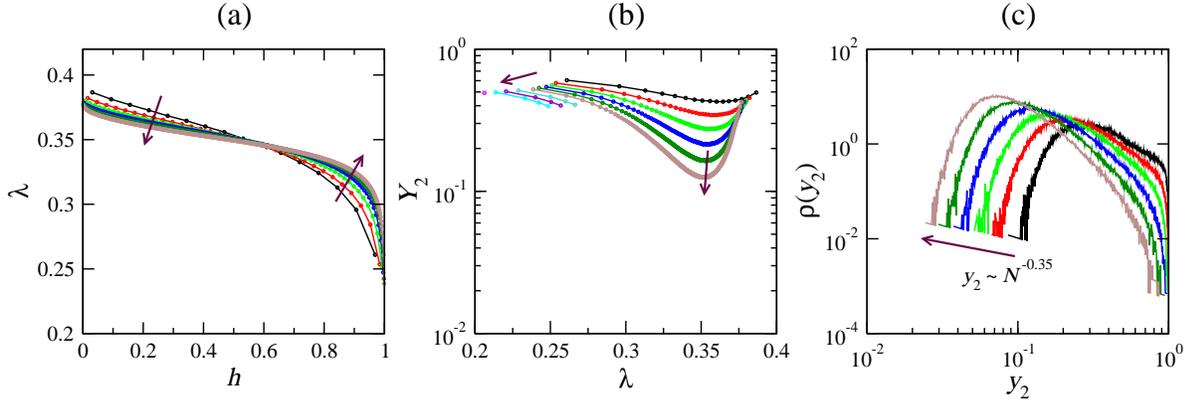}
  \caption{Lyapunov spectrum of the maps with finite $N$ for globally-coupled noisy logistic maps \eref{eq:GCMDef} with $a=1.86$, $K=0.10$, and $\sigma = 1 \times 10^{-3}$. (a) LE $\lambda^{(j)}$ vs rescaled index $h \equiv (j-0.5)/N$ and (b) parametric plots $(\lambda^{(j)}, Y_2^{(j)})$ for $N = 16, 32, \cdots, 512$ (increasing as indicated by the arrows). The last four modes are shown additionally for $N = 2048, 8192, 32768$ and the last mode is shown for $N = 131072$. (c) Distribution of the instantaneous IPR $y_2^{(j_{\rm min})}$ for the mode at the local minimum in the panel (b), $j_{\rm min} \equiv \arg\min Y_2^{(j)}$.}
  \label{fig:4-2}
% \end{center}
\end{figure}

\begin{figure}[t]
% \begin{center}
  \includegraphics[clip,width=\hsize]{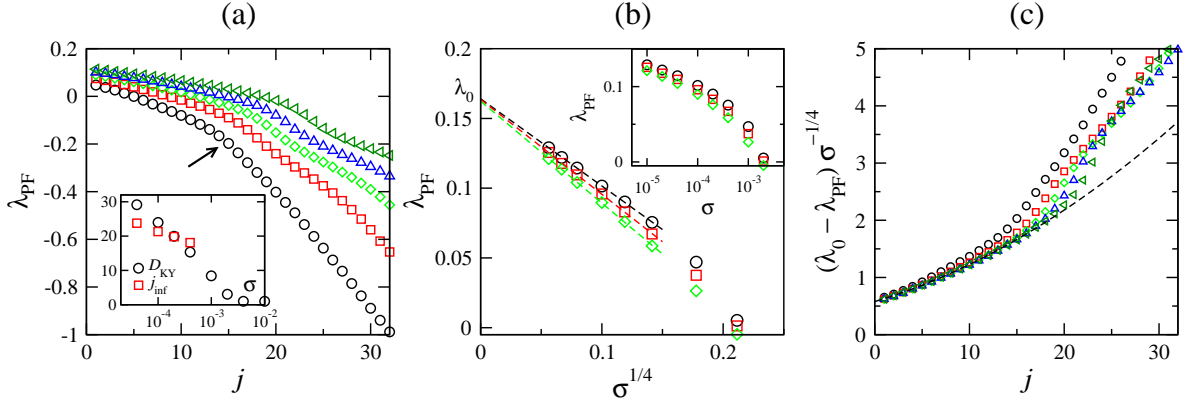}
  \caption{Lyapunov spectrum of the PF equation for globally-coupled noisy logistic maps \eref{eq:GCMDef} with $a=1.86$ and $K=0.10$. (a) PF Lyapunov spectrum $\lambda_{\rm PF}^{(j)}$ for various noise amplitudes
 $\sigma = 1 \times 10^{-3}, 4 \times 10^{-4}, 2 \times 10^{-4}, 1 \times 10^{-4}, 4 \times 10^{-5}$ (from bottom to top).
% $\sigma = 4 \times 10^{-5}, 1 \times 10^{-4}, 2 \times 10^{-4}, 4 \times 10^{-4}, \cdots, 1 \times 10^{-2}$ (from top to bottom). 
Sufficiently many bins are used in order that the shown values of the LEs are reliable.
% Note that only the first twelve LEs are shown for the three largest $\sigma$.
The arrow indicates a rough position of the threshold between the two regions, for $\sigma = 1 \times 10^{-3}$ (see text).
Inset: Kaplan-Yorke dimension $D_{\rm KY}$ and the index of the inflection point $j_{\rm inf}$ as functions of $\sigma$. (b) The first three PF LEs as functions of $\sigma$ (inset) and $\sigma^{1/4}$ (main panel). The dashed lines indicate the best linear fits to the data for $\sigma \leq 4 \times 10^{-4}$. (c) $(\lambda_0 - \lambda_{\rm PF}^{(j)}) \sigma^{-1/4}$ against $j$ (same symbols as in (a)) with $\lambda_0 = 0.163$.
% Data collapse found at small $j$ indicates that these LEs tend to the same asymptotic value $\lambda_0$ in the noiseless limit. 
The dashed line shows an estimate of the asymptotic curve $F_j$, obtained by a quadratic fit to the data for $\sigma = 4 \times 10^{-5}$ (green leftwards triangles) within $j \leq 17$.}
  \label{fig:4-3}
% \end{center}
\end{figure}

The Lyapunov spectrum of this system at $\sigma = 1 \times 10^{-3}$
 is measured both from the maps (Figure \ref{fig:4-2}(a,b))
 and from the PF equation (black circles in Figure \ref{fig:4-3}(a)).
One notices that, in this weak-coupling regime,
 even the largest LE in the PF dynamics,
 $\lambda_{\rm PF}^{(1)} \approx 0.047$, is much smaller
 than all the LEs in the maps with finite $N$ we investigated.
Indeed, none of the Lyapunov modes from the maps
% including the last, closest one to the PF modes,
 show signs of delocalization (Figure \ref{fig:4-2}(b)):
 the local minimum of $Y_2^{(j)}$ apparent in Figure \ref{fig:4-2}(b)
 depends only weakly on $N$, specifically $Y_2^{(j_{\rm min})} \sim N^{-0.35}$
 in contrast to $Y_2 \sim 1/N$ for the true delocalization,
 even in the distribution of the instantaneous IPR values $y_2$
 (Figure \ref{fig:4-2}(c); to be compared with Figure \ref{fig:2-4}(c)
 for the case of a true collective mode).
This local minimum is actually due to the near degeneracy of LEs
 around $\lambda^{(j)} \approx 0.34$ (Figure \ref{fig:4-2}(a)),
 which is typical of globally-coupled systems \cite{Takeuchi.etal-PRL2011}
 and thus also present in Figures \ref{fig:2-2} and \ref{fig:3-1}(a).
% and is not related to the collective modes.
Although the eventual emergence of collective modes is expected
 somewhere in the spectrum,
 which tends to cover a wider range of LE values for larger $N$,
 their detection seems out of reach of current computer power.
We therefore focus on the PF Lyapunov modes in the following,
 to investigate the nature of the non-trivial collective chaos.

The PF Lyapunov spectrum consists of gradually descending LEs,
% unless the noise amplitude is too large
 showing, roughly, two different slopes for the beginning and the rest
 of the spectrum, as indicated by the arrow in Figure \ref{fig:4-3}(a)
 for $\sigma = 1 \times 10^{-3}$.
With decreasing noise amplitude $\sigma$,
 both negative slopes increase
 and the threshold between the two regions shifts rightwards.
This obviously increases the value of the Kaplan-Yorke dimension $D_{\rm KY}$,
 which turns out to grow as $D_{\rm KY} \sim -\log\sigma$ for small $\sigma$
 (black circles in the inset)
 similarly to the result by Shibata \textit{et al.}
 \cite{Shibata.etal-PRL1999}.
In fact, the position of the threshold also moves logarithmically,
 as indicated by the index of the inflection point of the spectrum,
 $j_{\rm inf}$ (red squares in the inset).
In contrast, the values of the LEs $\lambda_{\rm PF}^{(j)}$
 with fixed indices cannot increase logarithmically,
 since they should remain finite in the noiseless limit.
Instead, our data suggest that at least first few LEs
 depend linearly on $\sigma^{1/4}$,
 i.e., $\lambda_{\rm PF}^{(j)} \simeq \lambda_0 - F_j \sigma^{1/4}$
 (Figure \ref{fig:4-3}(b)).
Moreover, the estimates of $\lambda_0$ from the fits to the first three LEs
 (dashed lines)
 coincide very well at $\lambda_0 = 0.163(1)$,
 where the number in the parentheses
 indicates a range of error in the last digit.
% for this asymptotic LE value.
Rescaling the Lyapunov spectra in Figure \ref{fig:4-3}(a)
 as $(\lambda_0 - \lambda_{\rm PF}^{(j)}) \sigma^{-1/4}$
 with the above estimate of $\lambda_0$,
 we find a reasonable collapse for small $j$ (Figure \ref{fig:4-3}(c)),
 which indicates that the asymptotic behavior
 $\lambda_{\rm PF}^{(j)} \simeq \lambda_0 - F_j \sigma^{1/4}$
 actually holds over the first region of the spectrum.
% towards the same asymptotic value $\lambda_0$.
This asymptotic form of the PF Lyapunov spectrum will be
 further investigated below.
%We will revisit this extrapolation on the asymptotic Lyapunov spectrum
% below.

\begin{figure}[t]
 \begin{center}
  \includegraphics[clip,width=.95\hsize]{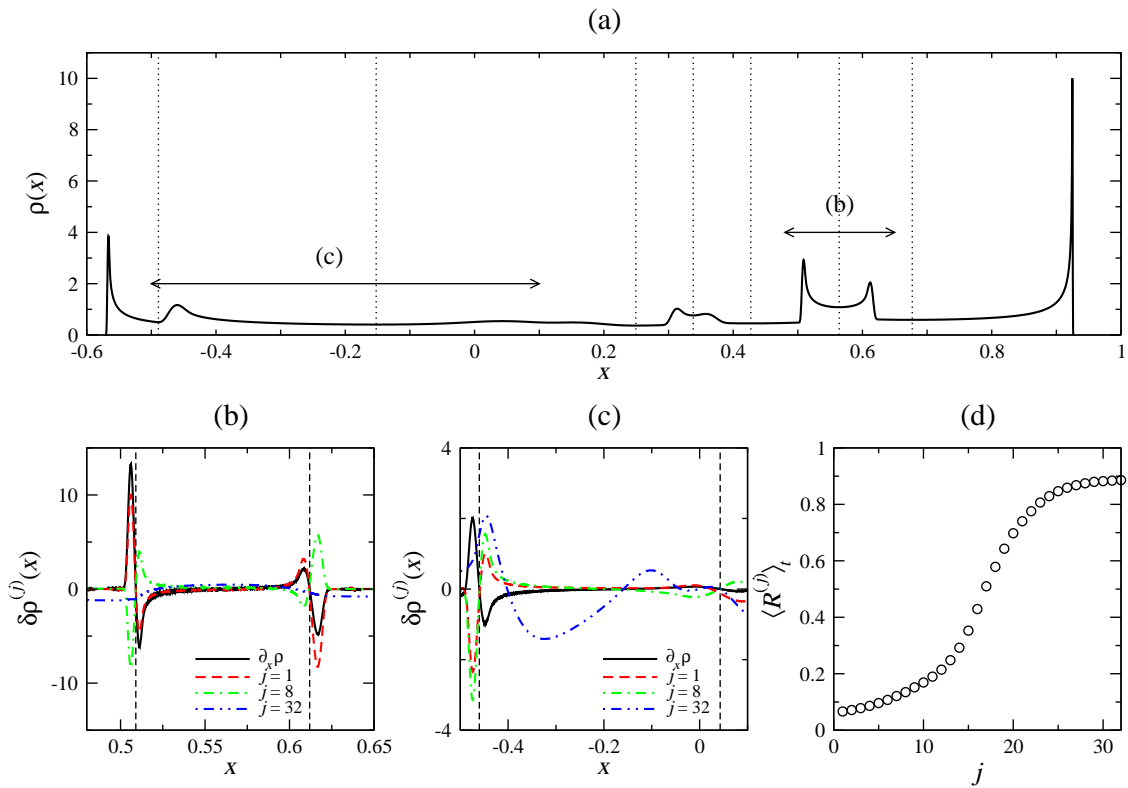}
  \caption{Snapshot of the distribution $\rho(x)$ (a) and CLVs $\delta\rho^{(j)}(x)$ (b,c) from the PF equation for globally-coupled noisy logistic maps \eref{eq:GCMDef} with $a=1.86$, $K=0.10$, and $\sigma = 1 \times 10^{-3}$. Vertical dotted lines in panel (a) show the partition of the support into the regions $S_k$, each containing one peak and bordered by local minima. In panels (b,c), black solid lines indicate the (arbitrarily scaled) first derivative of the distribution, the dashed lines show the CLVs $\delta\rho^{(j)}(x)$ with $j=1, 8, 32$ as stipulated in the legends, and the vertical dashed lines mark the positions of the peaks. Panel (d) shows the time-averaged residue $\expct{R^{(j)}}_t$ as a function of the index $j$ (see text).}
  \label{fig:4-4}
 \end{center}
\end{figure}%

\begin{figure}[t]
 \begin{center}
  \includegraphics[clip,width=.95\hsize]{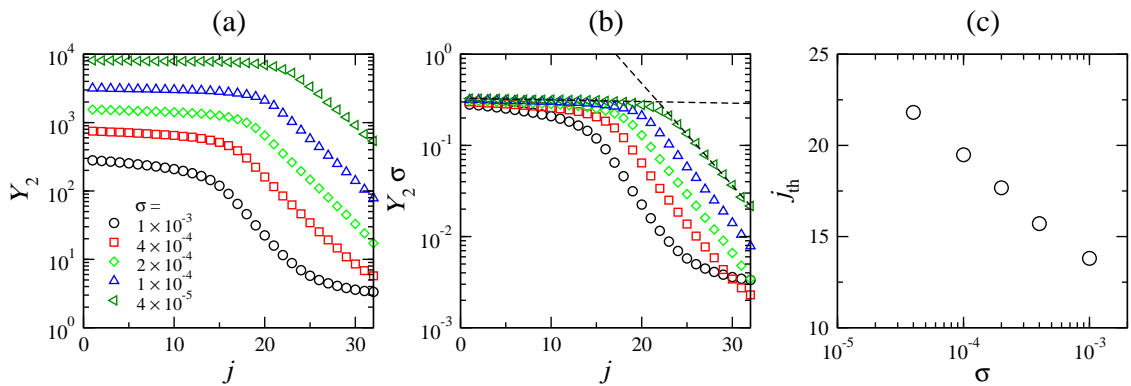}
  \caption{$Y_2^{(j)}$ spectrum of the PF Lyapunov modes for globally-coupled noisy logistic maps \eref{eq:GCMDef} with $a=1.86$ and $K=0.10$. (a) Spectra of the IPRs $Y_2^{(j)}$ for different noise amplitudes $\sigma$. (b) Same data are multiplied by $\sigma$. The dashed lines show linear fits (in the semilog scale) to the two spectrum regions for $\sigma = 4 \times 10^{-5}$ (top symbols), whose intersection determines the threshold index $j_{\rm th}(\sigma)$. (c) Threshold index $j_{\rm th}$ as a function of the noise amplitude $\sigma$.}%
  \label{fig:4-a}
 \end{center}
\end{figure}

We now turn our attention to the structure of the CLVs
 associated with these PF Lyapunov modes (Figure \ref{fig:4-4}).
Comparing the instantaneous distribution $\rho(x)$ (panel (a))
 and CLVs $\delta\rho^{(j)}(x)$ (panels (b,c)),
 we find that the CLVs in the first region, i.e. those with small $j$,
 tend to be concentrated near the peaks in the distribution
 (red dashed lines and green dotted-dashed lines in panels (b,c)).
Around each peak these CLVs are almost proportional to the first derivative
 of the distribution, $\rho'(x) \equiv \p_x\rho(x)$ (black solid line),
 but not globally, because the proportionality constant is different
 for each peak.
This indicates that these modes shift peaks in the distribution
 at different amplitudes and directions,
 similarly to the positive collective mode found
 in the limit-cycle oscillators (Figure \ref{fig:2-5}).
In contrast, CLVs with large $j$ do not resemble such a patchwork
 of local first derivatives but tend to be distributed widely,
 both around the peaks and on top of the broad plateaux in between
 (blue double-dotted-dashed lines in Figure \ref{fig:4-4}(b,c)).
They therefore deform the plateau structures and/or
 change the weight balance between the peaks and the plateaux.
To distinguish between these two types of Lyapunov modes,
 we quantify how similar each CLV $\delta\rho^{(j)}(x)$ is
 to the assembly of the local first derivatives,
 $\sum_k A_k^{(j)} \rho'_k(x)$,
 where $k$ denotes the region $S_k = [x_k, x_{k+1}]$ of the $k$th peak
 bordered at the local minima $x_k$ and $x_{k+1}$ of the distribution
 (dotted lines in Figure \ref{fig:4-4}(a))
 and $\rho'_k(x) = \rho'(x)$ if $x \in S_k$ and $0$ otherwise.
Specifically, we compute the residue
\begin{equation}
 R^{(j)} \equiv \sum_k \int_{S_k} \[ \delta\rho^{(j)}(x) - A_k^{(j)} \rho'(x) \]^2 \rd x  \label{eq:ResidueDef}
\end{equation}
 with the optimal choice of the coefficients,
% $A_k^{(j)}$, namely
 $A_k^{(j)} = \int_{S_k} \delta\rho^{(j)}(x) \rho'(x) \rd x / \int_{S_k} \rho'(x)^2 \rd x$.
This indeed shows a clear transition
% around $j \approx 17$ for $\sigma = 1 \times 10^{-3}$
 with varying $j$ (Figure \ref{fig:4-4}(d)),
 which underpins the existence of the two regions in the Lyapunov spectrum.
%Defining the threshold index $j_{\rm th}$
% by $\expct{R^{(j_{\rm th})}}_t = 0.5$,
% we find that it also depends logarithmically on the noise amplitude $\sigma$,
% as shown by red squares in Figure \ref{fig:4-5}(b).
The same transition can also be seen
 in the spectrum of the IPRs,
 defined here by $Y_2^{(j)} = \expct{\int \delta\rho^{(j)}(x)^4 \rd x}_t$
 with the normalization $\int \delta\rho^{(j)}(x)^2 \rd x = 1$
 (Figure \ref{fig:4-a}(a)).
Since the CLVs in the first region are localized around the density peaks,
 whose width is scaled by the noise amplitude $\sigma$,
% spectra of those $Y_2^{(j)}$ values
% overlap reasonably well onto a single flat line
 those IPRs take similar values
 when multiplied by $\sigma$ (Figure \ref{fig:4-a}(b)).
By contrast, IPRs in the second region
 decrease exponentially with increasing index $j$,
 setting a clear threshold index $j_{\rm th}$
%This allows us to define a threshold index $j_{\rm th}$
 for each noise amplitude $\sigma$
 (intersection of two linear fits in Figure \ref{fig:4-a}(b)).
The threshold index $j_{\rm th}$ is found to increase
 logarithmically with decreasing $\sigma$ (Figure \ref{fig:4-a}(c)),
 like $D_{\rm KY}$ and $j_{\rm inf}$
 shown in the inset of Figure \ref{fig:4-3}(a).

\begin{figure}[t]
 \begin{center}
  \includegraphics[clip,width=.95\hsize]{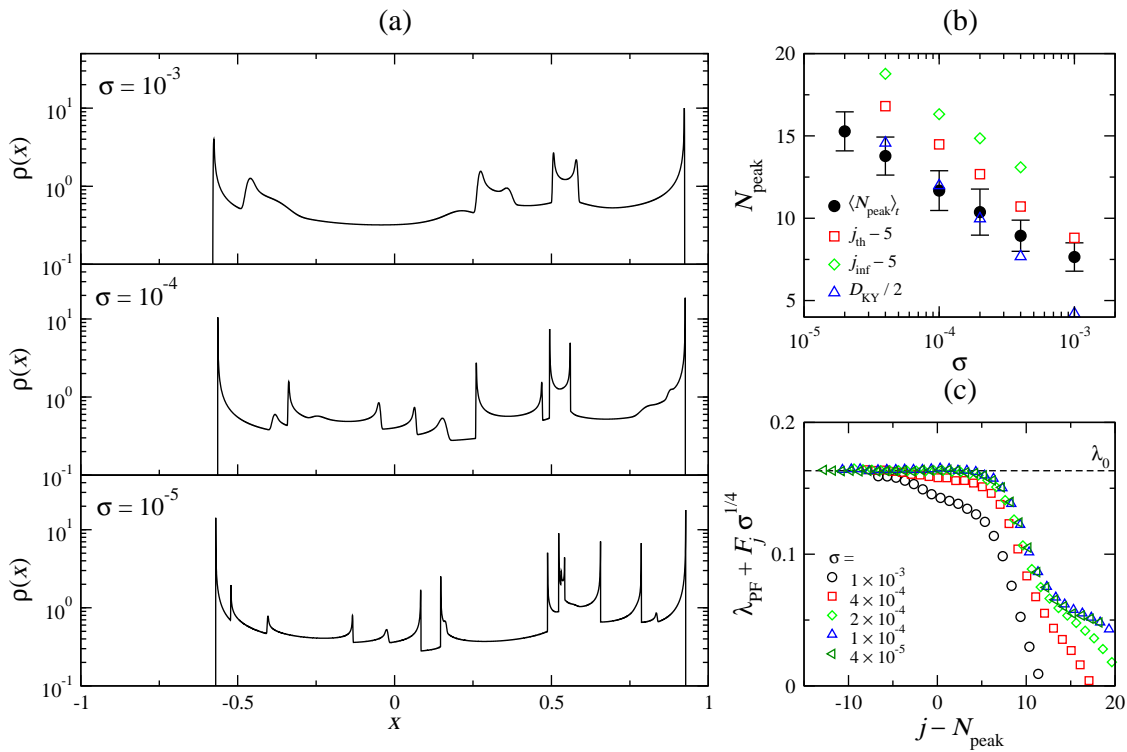}
  \caption{Number of density peaks and rescaling of the PF Lyapunov spectra for globally-coupled noisy logistic maps \eref{eq:GCMDef} with $a=1.86$ and $K=0.10$. (a) Typical instantaneous distributions $\rho(x)$ for different noise amplitudes $\sigma$. (b) Time-averaged number of peaks $N_{\rm peak}$ as a function of $\sigma$ (black solid circles). The bars indicate the standard deviation of the instantaneous values of $N_{\rm peak}$. The threshold index $j_{\rm th}$, the index of the inflection point $j_{\rm inf}$ and the Kaplan-Yorke dimension $D_{\rm KY}$ are also shown with shifts and factors stipulated in the legend. (c) PF Lyapunov spectra in the rescaled axes, exhibiting the asymptotic spectrum in the noiseless limit (same data and symbols as in Figure \ref{fig:4-3}(a)). The horizontal dashed line indicates our estimate of the asymptotic LE, $\lambda_0 = 0.163(1)$.
% (c) Spectra of the IPRs $Y_2^{(j)}$ in the original axes (inset) and the rescaled ones (main panel). The same symbols are used as in the panel (b).
}
  \label{fig:4-5}
 \end{center}
\end{figure}%

The role played by the Lyapunov modes in the first region
 against the density peaks in the distribution
suggests a direct connection between their numbers.
Indeed, decreasing the noise amplitude $\sigma$,
 we find more and more peaks in the instantaneous distributions
 (Figure \ref{fig:4-5}(a)), simply because
 the sharpest peak created from the superstable point
 becomes sharper and sharper.
We then count
% and plot in Figure \ref{fig:4-5}(a)
 the time-averaged number of peaks $N_{\rm peak}$
 for each noise amplitude $\sigma$,
 and find that it varies logarithmically
% with $\sigma$
 as $N_{\rm peak} \simeq N_0 - C_{\rm peak}\log\sigma$
 (black solid circles in Figure \ref{fig:4-5}(b)),
% (black solid circles)
% and the number of the Lyapunov modes in the first region (red solid squares),
% which is estimated by the threshold index $j_{\rm th}$
% positioned at $\expct{R^{(j_{\rm th})}}_t = 0.5$.
%We then find that, indeed,
 similarly to the number of the Lyapunov modes in the first region,
 or the threshold index, $j_{\rm th} \simeq j_0 - C_{\rm th}\log\sigma$
 (red squares; reproduced from Figure \ref{fig:4-a}(c)).
Moreover, the two coefficients are found to be very close,
 i.e., $C_{\rm peak} \approx C_{\rm th}$
 (same slope in Figure \ref{fig:4-5}(b)).
This indicates that, 
% as the number of peaks increases by one
% with decreasing noise amplitude $\sigma$,
 for each peak added by decreasing the noise amplitude $\sigma$,
 a new Lyapunov mode is introduced to the first region of the PF spectrum,
 in order for this mode and the other existing ones
 to describe the instability of the newly added peak.
The number of the peaks $N_{\rm peak}$ therefore controls
 the form of the Lyapunov spectrum in the first region:
 using $\lambda_{\rm PF}^{(j)} \simeq \lambda_0 - F_j \sigma^{1/4}$
 reported in Figure \ref{fig:4-3}(b,c),
 we find a reasonably good collapse of the PF Lyapunov spectra
 within the axes $\lambda_{\rm PF}^{(j)} + F_j \sigma^{1/4}$
 vs $j - N_{\rm peak}$ for sufficiently small noise amplitudes
 (Figure \ref{fig:4-5}(c)).
In particular,
 the ordinates in this representation indicate
 the LE values in the noiseless limit $\sigma \to 0$,
 which are surprisingly flat at $\lambda_0 = 0.163(1)$
 in the first region.
%Similarly, the spectrum of the IPRs,
% defined here by $Y_2^{(j)} = \int \delta\rho^{(j)}(x)^4 \rd x$
% with the normalization $\int \delta\rho^{(j)}(x)^2 \rd x = 1$,
% can also be rescaled with $\sigma$ and $N_{\rm peak}$.
%For the three smallest $\sigma$, 
% the raw data in the inset of Figure \ref{fig:4-5}(c) are collapsed
% onto a single curve in the main panel
% by replacing $j$ with $j-N_{\rm peak}$
% and multiplying $Y_2^{(j)}$ by $\sigma$ (main panel),
% since the CLVs for the first region are localized around the peaks,
% whose widths are simply proportional to the noise amplitude $\sigma$.

To sum up, using the CLVs and the controlled noise distribution,
 we arrive at the following conclusion reached by Shibata \textit{et al.}
 \cite{Shibata.etal-PRL1999} on a firm basis:
 the effective dimension $D$ of the non-trivial collective chaos
 is finite when noise is added to the system,
 but it increases logarithmically with decreasing noise amplitude,
 $D \sim -\log\sigma$, and in particular diverges in the noiseless limit.
In other words, the collective chaos in the noiseless system
 has infinite dimensionality.
%Using the PF evolution, 
We have shown that this logarithmic divergence is
 due to the increasing number of the leading Lyapunov modes
 in the PF description, which exert various translational shifts
 to the density peaks in the distribution of the dynamical units.
Therefore, the effective dimension of the collective chaos
 can be estimated, at least in this regime,
 by the number of these peaks $N_{\rm peak}$
 as demonstrated in Figure \ref{fig:4-5}(a),
 which can also be measured from the microscopic simulations of the maps,
 in principle.

\begin{figure}[t]
 \begin{center}
  \includegraphics[clip,width=.95\hsize]{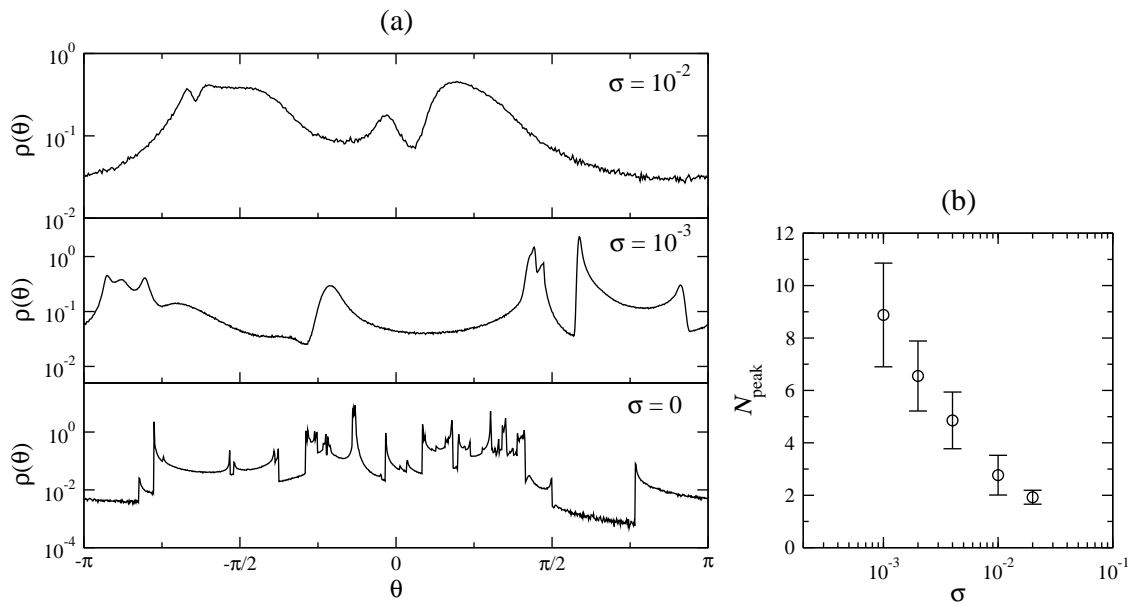}
  \caption{Instantaneous distributions (a) and number of peaks $N_{\rm peak}$ (b) for globally-coupled noisy limit-cycle oscillators \eref{eq:NoisyGLDef} with various noise amplitudes $\sigma$. Configurations without folding are chosen for the instantaneous distributions shown in the panel (a), which are projected onto the angular coordinates $\theta_i \equiv \arg W_i$. The distribution for $\sigma = 0$ is recaptured from Figure \ref{fig:2-1}(b). For panel (b), symbols and bars indicate the mean and the standard deviation of the instantaneous numbers of the peaks. For both panels, the numbers of the oscillators are chosen between $10^6$ and $10^8$ to obtain reliable results.}
  \label{fig:4-6}
 \end{center}
\end{figure}%

On the basis of these results obtained
 for globally-coupled noisy logistic maps,
 we finally revisit the collective chaos studied in Section \ref{sec:Col},
 exhibited by the globally-coupled limit-cycle oscillators.
Given the similar structure of the instantaneous distributions
 (compare Figures \ref{fig:2-1}(b) and \ref{fig:4-5}(a))
 and the similar role of the collective or PF Lyapunov modes
 (Figures \ref{fig:2-5}(a,b) and \ref{fig:4-4}),
 we expect that these two systems may share
 basic properties of the collective chaos.
We therefore consider the limit-cycle oscillators in the presence of noise,
 added here sporadically for the sake of simplicity:
\begin{equation}
 \dot{W}_i = W_i - (1+\mathrm{i}c_2)|W_i|^2 W_i + K (1+\mathrm{i}c_1) (\expct{W} - W_i) + \sum_{n\in\mathbb{Z}} \delta(t - n)\xi_i^n,   \label{eq:NoisyGLDef}
\end{equation}
 where $\xi_i^n$ is drawn independently from the uniform distribution
 $[-\sigma,\sigma]$.
We then measure the oscillator density projected on the angular coordinates
 $\theta_i \equiv \arg W_i$ for various noise amplitudes $\sigma$,
 and indeed find more peaks for lower values of $\sigma$
 (Figure \ref{fig:4-6}(a)).
Counting the number of the peaks therein, averaged along the trajectory,
 we again identify a logarithmic increase $N_{\rm peak} \sim -\log\sigma$
 (Figure \ref{fig:4-6}(b)) like in the case of the logistic maps.
Therefore, if we assume that there are
 roughly as many Lyapunov modes as the peaks for shifting them
 in this system too,
 the logarithmic growth in $N_{\rm peak}$ implies
 that the effective dimension of the collective chaos
 also increases and diverges in the noiseless limit.
In this case, one should find more and more collective Lyapunov modes
 in addition to the ones found in Section \ref{sec:Col}
 as increasing the system size further, though it is unattainable
 with the current machine power.
Similarly, one can in principle study Lyapunov modes
 associated with the PF evolution for this system,
 but it seems to be unfeasible to track
 the evolution of the intricate, fractal-like density profile $\rho(W)$
 in the complex plane (see Figure \ref{fig:2-1}(c))
 in a reliable manner with a finite number of bins.
Although our results suggest that the logarithmic divergence takes place
 in $N_{\rm peak}$ as well as in the dimension of the collective chaos,
 all the more because peaks are formed and dispersed
 by stretching and folding of the support (and noise, if any)
 similarly to the logistic maps,
 providing a direct evidence for it, either numerically or theoretically,
 is a challenging open problem left for future studies.
%Such numerical attempts,
% as well as theoretical approaches to this issue,
% are left as challenging open problems.

\section{Discussions and concluding remarks}  \label{sec:Dis}

In this paper, we have first shown that
 the standard Lyapunov analysis does capture
 the collective dynamics of large chaotic systems:
 it is encoded in a set of collective Lyapunov modes,
 which are delocalized over the collection of the dynamical units
 and thus exert relevant perturbations to macroscopic variables,
 without the need for finite-amplitude perturbations
 as opposed to some earlier claims
 \cite{Shibata.Kaneko-PRL1998,Cencini.etal-PD1999}.
The CLVs allow us to detect such collective modes
% with a clear and quantitative criterion for the delocalization,
 with the delocalization criterion, $Y_2^{(j)} \sim 1/N$,
 and to examine directly their role in the collective dynamics,
 as demonstrated for the regime of non-trivial collective chaos
 in globally-coupled limit-cycle oscillators (Section \ref{sec:Col}).
Moreover, for globally-coupled systems,
 the instabilities of the collective dynamics can also be studied
 by the associated PF equation,
 whose Lyapunov modes, at least some of them, coincide
 with the collective modes of the original dynamical systems
 (Section \ref{sec:ColPF}).
On the basis of this correspondence,
 in Section \ref{sec:ColChaos},
 we have analyzed in detail the non-trivial collective chaos
 in globally-coupled noisy logistic maps.
We have then found leading PF Lyapunov modes,
 thus leading collective modes,
 assigned to translational shifts of dense clusters
 loosely formed by the dynamical units.
The number of such clusters, $N_{\rm peak}$ increases logarithmically
 with decreasing noise amplitude $\sigma$,
 i.e., $N_{\rm peak} \sim -\log\sigma$,
 and so do the number of leading collective modes
 and other effective dimensions of collective chaos.
This is expected to be a common feature of collective chaos
 that takes place on a bounded support undergoing stretching and folding,
 and indeed the logarithmic increase of $N_{\rm peak}$ was found
also for globally-coupled noisy limit-cycle oscillators.

\begin{figure}[t]
 \begin{center}
  \includegraphics[clip,width=.4\hsize]{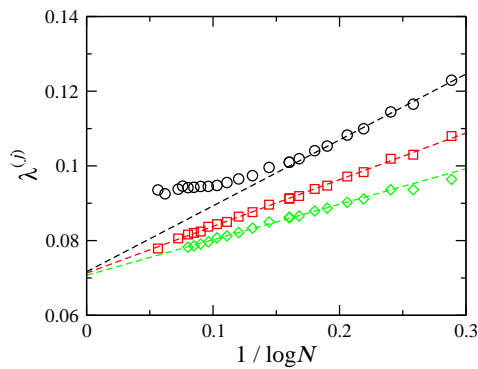}
  \caption{System-size dependence of the first three LEs for globally-coupled limit-cycle oscillators \eref{eq:GLDef}. The data shown in Figure \ref{fig:2-4}(b) are plotted against $1/\log N$. The dashed lines show linear fits to the $1/\log N$ regime.}
  \label{fig:5-1}
 \end{center}
\end{figure}%

The existence of the collective modes,
 whether their number is finite or slowly increasing with system size,
 calls for revisiting the conventional definition of the extensivity of chaos
 in the presence of collective behavior,
 which assumes all the Lyapunov modes in the spectrum to be extensive.
For globally-coupled systems, this form of extensivity
 does not hold even in the absence of collective behavior:
 in this case extensive LEs are sandwiched by ``subextensive'' bands
 composed of $\mathcal{O}(\log N)$ LEs at both ends of the spectrum,
 which vary logarithmically with system size
 as $\lambda \sim \lambda_\infty + \const/\log N$
 \cite{Takeuchi.etal-PRL2011}.
This picture is modified when collective behavior takes place
 (Figure \ref{fig:5-1}):
 taking the first few LEs
 for the limit-cycle oscillators shown in Figure \ref{fig:2-4}(b)
 and representing them as functions of $1/\log N$,
 we find that the second and subsequent LEs indeed obey
 the logarithmic size dependence for the subextensive exponents
 (dashed lines),
 while the first, collective mode (black circles) deviates from it
 as it develops the delocalized CLV structure with increasing system size.
Moreover, as we have discussed in the previous section,
 there could exist more and more collective modes
 as the system size $N$ is further increased.
Recalling the fact that the global field admits
 statistical fluctuations proportional to $1/\sqrt{N}$
 around its time-evolving mean value \cite{Pikovsky.Kurths-PRL1994},
 we might consider that each dynamical unit is subjected to
 an effective noise of amplitude $\sigma \sim 1/\sqrt{N}$ in this case.
Then, our finding on the number of the density peaks implies that
 the number of the collective modes increases as $-\log\sigma \sim \log N$,
 adding another logarithmic subextensive band to the Lyapunov spectrum.
A theoretical study is certainly needed to put
 this speculation on a firm basis.
Seeking for a direct evidence of a collective mode
 in locally coupled systems is also an important issue
 left for future studies, in particular for the general understanding
 of the chaos extensivity.

Let us finally turn to the often-discussed but never-demonstrated analogy
 of NTCB to small dynamical systems.
% effect of the varying control parameters on the collective Lyapunov modes.
We have found for the limit-cycle oscillators that the collective LEs
 have the same set of the signs as what we would expect
 from the observed macroscopic dynamics,
 whether the number of positive collective LEs increases or not,
 and hence underpinned the analogy of NTCB to small dynamical systems.
There is, however, an essential difference here:
 in contrast to small systems,
 the total number of the collective modes is not determined \textit{a priori}.
This implies that large dynamical systems may have far richer
 bifurcation structure at the macroscopic level,
 where collective exponents may not only change their signs
 but also be created and annihilated
 at the transition between two different NTCB regimes.
It is important to clarify how this viewpoint is reconciled
 with some features like phase transitions, such as critical phenomena,
 reported by a few earlier studies
 \cite{Lemaitre.Chate-PRL1999,Marcq.etal-PTPS2006}.
The collective Lyapunov modes should play central roles
 in tackling these fundamental issues left hitherto unexplored,
 which we believe deserve rather high computational cost
 needed for the investigations.

\section*{References}
\bibliographystyle{iopart-num}
\bibliography{JPAcollective,otherrefs}

\end{document}